# Astrometry with PRAIA

M. Assafin[a,b]

[a]*Universidade Federal do Rio de Janeiro - Observatório do Valongo, Ladeira do Pedro Antônio 43, Rio de Janeiro, 20.080-090, RJ, Brazil*
[b]*Laboratório Interinstitucional de e-Astronomia (LIneA), Rua General José Cristino 77, Rio de Janeiro, 20.921-400, RJ, Brazil*

**Abstract**

`PRAIA` – Package for the Reduction of Astronomical Images Automatically – is a suite of astrometric and photometric tasks designed to cope with huge amounts of heterogeneous observations with fast processing, no human intervention, minimum parametrization and yet maximum possible accuracy and precision. It is the main tool used to analyse astronomical observations by an international collaboration involving Brazilian, French and Spanish researchers under the Lucky Star umbrella for Solar System studies. In this paper, we focus on the astrometric concepts underneath `PRAIA`, used in reference system works, natural satellite and NEA astrometry for dynamical and ephemeris studies, and lately for the precise prediction of stellar occultations by planetary satellites, dwarf-planets, TNOs, Centaurs and Trojan asteroids. We highlight novelties developed by us and never reported before in the literature, which significantly enhance astrometry precision and automation. Such as the robust object detection and aperture characterization (BOIA), which explains the long standing empirical photometry/astrometry axiom that recommends using apertures with $2-3\ \sigma$ (Gaussian width) radius. We give examples showing the astrometry performance, discuss the advantages of `PRAIA` over other astrometry packages and comment about future planed astrometry implementations. `PRAIA` codes and input files are publicly available for the first time at: `https://ov.ufrj.br/en/PRAIA/`. `PRAIA` astrometry is useful for Solar System as well as astrophysical observations.

## 1. Introduction

Astrometry of digitized images from ground-based telescopes was very useful in the conciliation of optical and radio reference systems in the 90's and 2000's (Assafin et al., 2013). It is also very important in Solar System dynamics and ephemeris studies (Camargo et al., 2022). Nowadays it is fundamental in the prediction of stellar occultations by Solar System bodies such as planets, natural and irregular satellites, main belt asteroids, Near Earth and Potentially Hazardous Asteroids (NEAs, PHAs), trojan objects, Centaurs, transneptunian objects (TNOs), and dwarf planets (Desmars et al., 2015). Stellar occultations allow for the precise determination of the sizes and shapes of these bodies, and can also furnish valuable information on putative atmospheres, the detection of faint rings, moons and disperse material around them (Morgado et al., 2023).

A small group of Brazilian researches led a number of pioneering astrometric works in reference systems and Solar System bodies since the 90's (see Table 4, Section 11). It has grown up and since the mid-2000's the now known as Rio Group started to lead and participate on many important studies of Solar System bodies by using stellar occultations, giving fundamental astrometric and photometric support (see Table 4). The Rio Group maintains a long term international collaboration in the Lucky Star Project[1] with two other teams, the Meudon/Paris (France) and Granada (Spain) groups[2]. This collaboration involves the use of many small (50 cm) to mid (1–2 m) to large (4–8 m) aperture telescopes in Chile, Hawaii, Spain, France, Brazil, Australia and Africa, with many types of CCD detectors. The collaboration maintains a fruitful interaction with the world amateur community which actively participates on our stellar occultation campaigns. The amateur participation is most important as it covers many events that would otherwise be missed. In this case, telescope apertures range from 20 cm to 1 m, and observations may be done under non-ideal sky/technical conditions with poor guiding, low signal-to-noise ratios (S/N), few calibration stars, timing issues, etc.

Reference system astrometric observations were always voluminous due to dithering to overcome the limited sizes of the field of view (FOV) and sparse available reference catalogue stars in the 90's and 2000's. CCD observations for improving the dynamics/ephemeris of Solar System bodies and for the precise prediction of stellar occultations also imply in large amounts of images (10–50 GB per run), with 1–600 s exposures in a balance between S/N and apparent motion. The astrometric observations from the Rio Group and Lucky Star form a heterogeneous data set acquired with many telescope types and detectors, under a wide range of sky conditions and S/N regimes.

We soon realized that it was not practical to adapt for astrometry some standard packages available in the 90's such as `SExtractor` (Bertin & Arnouts, 1996), the Image Reduction and Analysis Facility (`IRAF`[3], Tody 1986, 1993) and `DAOPHOT` (Stetson, 1987), due to the excessive parameterization and human interaction. The solution was to develop our own pipeline for the astrometry of digitized images, and the PRAIA astrometry task (PAT) was born. PAT is meant at full automation, minimum parameterization, fast processing speed and high precision. It does not call external packages. PAT was used on 53 works published between 1992–2002 in prestigious journals. Here, we describe the astrometric concepts underneath PAT, highlighting implemented astrometric innovations. We are releasing PAT for the public for the first time.

---

[1]Lucky Star: `https://lesia.obspm.fr/lucky-star/index.php`
[2]Lucky Star team: `https://lesia.obspm.fr/lucky-star/team.php`

[3]`IRAF` was distributed by the National Optical Astronomy Observatory,

In Section 2 we overview the task evolution, current functionality and concepts, highlighting the novelties and last upgrades, such as the fully automatic object detection and analysis BOIA (Browsing Objects: Identification and Analysis), and the new $(x, y)$ measuring method PGM (Photogravity Center Method) that improves the Modified-Moment algorithm and furnishes $(x, y)$ error estimates, among others. BOIA is described in Section 3. In Section 4 we present the $(x, y)$ centring methods: circular and elliptical Gaussian and Lorentzian (extended Moffat) profiles, and PGM. Catalogues are described in Section 5 and target data (ephemeris, etc) in Section 6. Position reduction to observation date is shown in Section 7. Details about the $(x, y)$ to $(\alpha, \delta)$ reduction are given in Section 8, including the fully automatic identification of reference catalogue stars, and the two-step $(\alpha, \delta)$ reductions for improving measurements and 100 per cent recovery of missing targets and reference catalogue stars. Output of positions, Point-Spread-Function (PSF) magnitudes, errors and FOV visualization of results are commented in Section 9. In Section 10 we validate PAT by a thorough error analysis of measurements from 5 distinct types of FOVs common in our Solar System work, with typical CCD astrometry problems. We show that the S/N optimization – core of BOIA procedures – explains (and extends) the old astrometry/photometry axiom that dictates the best relation between aperture sizes and the Full-Width-Half-Maximum (FWHM) of objects. Section 11 also certifies PAT with the report of 30 years of referred works published in prestigious journals, that used the task for the main results, and for full or partial astrometric support of results. Supplementary PRAIA astrometry tasks are presented in Section 12: an ephemeris extraction task and novel astrometric methods developed by us – the Mutual Approximations and Differential Distance Astrometry, respectively inspired in mutual phenomena and differential photometry techniques. Section 13 discusses the characteristics of old/new popular astrometric packages and PAT in the context of the Rio Group and the Lucky Star collaboration. Conclusions, remarks and future task updates are drawn in Section 14.

## 2. The PRAIA astrometry task – an overview

PRAIA stands for "Package for the Reduction of Astronomical Images Automatically". It is a standalone package fully developed in FORTRAN with no graphic interfaces, composed of multiple independent tasks. Each task has a source/executable code and an input text file (ASCII format) with editable input parameters. The tasks can be arranged in scripts for the reduction of astrometric and photometric observations in (standard or not) Flexible Image Transport System (FITS, Wells et al. 1991) format. PRAIA gives full control for the user to manipulate input and output data with external statistical, plotting and visualization tools. Differential aperture photometry with PRAIA is presented in Assafin (2023a). Digital coronagraphy, another important PRAIA tool used both in photometric and astrometric works, is described in Assafin (2023b).

The PRAIA astrometry task (PAT) is fundamental to the Rio Group and Lucky Star collaboration. It has been used in the study of reference systems, natural satellite and NEA astrometry for dynamics and ephemeris works, and lately for the precise prediction of stellar occultations by planetary satellites, dwarf-planets, TNOs, Centaurs and Trojan asteroids. Detailed installation/usage instructions for the task, code, input files and documentation are for the first time publicly available[4]. Here, we give the concepts of PAT, highlighting novelties introduced for the first time in astrometry. PAT automatically performs fast, accurate and precise small FOV tangent plane astrometry on vast numbers of images. It also does astrometry from lists of $(x, y)$ or $(\alpha, \delta)$ measurements. $(\alpha, \delta)$s, errors and other astrometric and photometric data are output. PAT is suited for the astrometry of point-like to moderately extended sources like stars, quasars, AGNs, NEAs, asteroids, Centaurs, TNOs, natural and irregular satellites, planets, dwarf-planets and comets.

PAT is in its fourth version. Version 0 (1990–2000) did semi-automatic astrometry of digitized photographic plates and CCD images. Astrometric catalogues were accessed from local hard disks. The $(x, y)$s came from 1D Gaussian fittings. Identification of reference catalogue stars was not automatic. We used old versions of ds9[5] as an auxiliary tool. Version 1 (2000–2007) introduced major changes making PAT fully automatic. The sky background of the FOV was flatted by bi-variate $(x, y)$ polynomials and the threshold for object detection was obtained by histogram analysis (Bijaoui, 1980). The 2D Gaussian was adopted. Pixel scale and error were needed for the identification of reference stars by cross-matching between measured/catalogue star pairs. Version 2 (2007–2019) introduced a new procedure for object detection that mitigated the effects of sky background variations. The FOV was recursively searched from brightest to fainter grouped pixels, avoiding the need to set subjective FOV-location-dependent factors for sky background statistics. Detections were fitted by a 2D Gaussian within 1 FWHM from the centre. Old task releases prior to 2000 derived very good results – the available reference catalogues lacked parallaxes, and proper motions were poorly known or absent. Disk space was not an issue as the catalogues were not large. But with the outcome of the Hipparcos mission (ESA, 1997) and the adoption of the new International Celestial Reference System (ICRS, Feissel & Mignard 1998), more precise and star-abundant astrometric

---

which is operated by the Association of Universities for Research in Astronomy (AURA) under cooperative agreement with the National Science Foundation. Currently it is supported by the IRAF Community: https://iraf-community.github.io

[4] PAT is publicly available at https://ov.ufrj.br/en/PRAIA/
[5] ds9 (SAOImageDS9): http://ds9.si.edu/site/Home.html

catalogues started to become available in the 2000's for small FOV astrometry, such as the UCAC catalogue (Zacharias et al., 2000). The demand for higher astrometric precision started to grow. The old procedures were no longer satisfactory in certain cases. The $(x, y)$ measurement of slightly elongated images (guiding issues, wind, etc) were not optimal. Depending on the observation conditions (Moon, close-by bright objects, vignetting, seeing variation, etc), the sky background and the seeing could significantly vary within the FOV or from one image to the other, challenging the setup of parameters for the automatic detection of objects. Crowded star FOVs and large ones imposed difficulties on identifying reference catalogue stars and targets. Needs for task improvements become imperative with the Gaia DR1 catalogue release in 2016 (Gaia Collaboration, 2016). Although lacking parallaxes and proper motions, Gaia DR1 had much better accuracy and precision than the UCAC4 catalogue (Zacharias et al., 2013). The Gaia DR1 was incorporated to the task version 2 with the hybrid UCAC5 – a mixture of Gaia DR1 and UCAC4 with proper motions (Zacharias et al., 2017). Local hard disk storage of Gaia DR1 and UCAC5 catalogues was becoming prohibitive, with loss of speed access to their content too. Then the Gaia DR2 catalogue was released in 2018 (Gaia Collaboration, 2018a,b), followed by the Gaia eDR3 catalogue in 2021 (Gaia Collaboration, 2021) and the Gaia DR3 in 2022 (Gaia Collaboration, 2022) with 1.5 billion stars up to magnitude 20.5 with positions, annual proper motions and astrometric parallaxes precise to better than 1 mas.

Major astrometry upgrades/novelties were finally introduced in the last PAT version 3 (2019) from object detection to $(x, y)$ measurements to $(\alpha, \delta)$ reductions. Astrometric improvements came together with photometric ones, also implemented in the last version of the PRAIA photometry task (Assafin, 2023a). Both tasks now share common and novel procedures. We tied fundamental photometry and astrometry concepts allowing for their mutual improvement. Now PAT accesses catalogues over the internet. Gaia DR3 is the main task catalogue representing the ICRS in the optical domain. We can select the magnitude range of reference objects and eliminate Gaia stars lacking annual proper motions, parallax, radial velocity or having star duplicity flags. We extract infrared magnitudes from the 2MASS catalogue (Cutri et al., 2003). Using our original BOIA procedure (Browsing Objects: Identification and Analysis, Section 3) – also used in the updated photometry and digital coronagraphy PRAIA tasks – we now automatically detect objects in the FOV without any sky background parameters. Widespreadly used in astrometric/photometric programs, sky background factors were completely abolished with BOIA. Procedures now eliminate false-positive detections like cosmic rays, diffraction-spike artifacts, etc. Masks for the elimination of bad pixels are now available for circular and rectangular regions, and ADU (Analogic Digital Unit) count ranges. Many PSF models are now available for the $(x, y)$ centring: 2D circular and elliptical Gaussian and Lorentzian (extended Moffat) profiles, with elimination of pixel outliers in the PSF fittings. We also introduced a novel 2D modified moment centring algorithm named Photogravity Center Method (PGM) in analogy to the gravity centre of a distribution of particles. The $(x, y)$ to $(\alpha, \delta)$ reduction is made in apparent coordinates with full models (parallax, annual proper motions, radial velocities) for better tackling meteorological conditions, refraction and bandpass of observations. Reference star position weights from $(x, y)$ errors are now available. Elimination of outliers is by sigma-clipping or by |O-C| cutoff in $(\alpha, \delta)$ reductions (details in the User Guide). Solar System objects have output positions corrected by solar phase angle. Two $(\alpha, \delta)$ reductions allow the identification of bad-measured/missing reference stars and targets, improving the astrometry of all objects. Richer ds9 region files are now output, improving visual experience.

## 3. Automatic object identification and analysis

### 3.1. BOIA (Browsing Objects: Identification and Analysis)

BOIA (Browsing Objects: Identification and Analysis) is an ensemble of original procedures developed by us, aiming at the robust automatic detection, optimal aperture setup and characterization of the image properties of objects in the FOV. BOIA is present in parts or in totality on all astrometry/photometry/coronagraphy PRAIA tasks. BOIA uses original fast procedures for object identification fundamentally different from the common algorithms of other packages usually based in SExtractor (Bertin & Arnouts, 1996). The setup of these packages is complex, depends on subjective factors related to the local or global sky background and sometimes user intervention is needed in running time (see Section 13). This contrasts with BOIA: no sky background factors are necessary, no user intervention, object identification and analysis is fully automatic.

Section 3.2 shows the quartile statistics used to improve sky background and ring computations. The core of the original BOIA's detection procedures are presented in Sections 3.3 and 3.4. Primary detections are described in Section 3.3. The determination of optimized apertures and sky background rings characterizing detections is detailed in Section 3.4. There may be multiple detections per object at the end of this stage, including false-positive detections. The selection of the best final detection per true object – i.e. the object identification itself – and the elimination of false-positive ones are described in Section 3.5. The image properties of the identified objects are obtained as in Section 3.6. They consist of the aperture centre, central moments, semi-major and semi-minor axes and rotation angle. Image properties are used as input data in the PSF fittings for the determination of the $(x, y)$ centre of the objects (Section 4.3). Missing detections of known objects (reference catalogue objects, targets) are recovered after a primary $(x, y)$ to $(\alpha, \delta)$ reduction. The procedure for setting optimum aperture parameters for missing detections are shown in Section 8.4.

### 3.2. Quartile statistics on rings and sky background with BOIA

Histogram statistics on rings in the inner regions of the object are specially affected. Less pixels are available as compared to the outer regions, resulting in poor histogram statistics and biased mode counts. In quartile statistics, the smallest and highest quarter values of the data are discarded, and the 50 per cent middle values used to estimate the average and standard deviation of the sample. The extremes may contain corrupted information, while the middle always provides better statistics for estimating the basic properties of the data sample. By using quartile statistics, we improved the computations on rings in substitution to histograms used in older task versions.

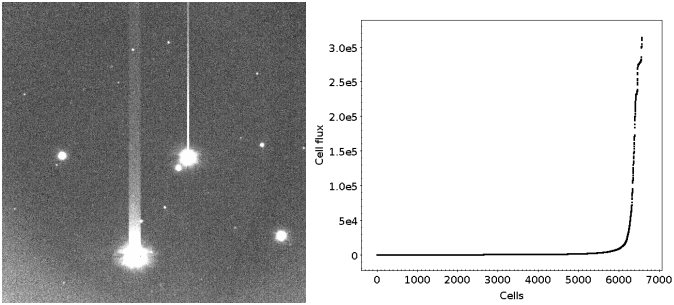

Figure 1: Surviving cells distributed in ascending flux order (right) for a sampled FOV (left). Higher fluxes belong to parts of bright or well-sampled objects, while low fluxes are associated to their wings and to the central parts of fainter objects. Many distinct cells with varying fluxes may be part of the same true object. In the example, 6577 surviving cells were found.

For the sky background computations using a passing median filter (see Section 3.3), the quartile limits are refined since sometimes the sampled pixels may contain large portions of objects. The number of middle pixels is still 50 per cent. After sampling all pixel counts in ascending order, we sample all possible lower/upper bounds, keeping the values with the smaller gradient, i.e. the closer lower/upper bound values, which set the best representatives of the sky background sampled pixels. Thus, values other than 25/25 per cent lower/upper bounds may occur for the fainter/brighter quartiles (e.g. 15/35, 20/30 per cent, etc), resulting on better elimination of flux contamination in the sky background computations.

### 3.3. Object detections with BOIA

We first pass a median filter for estimating the local sky background over the whole FOV, to set local thresholds for eliminating pixels not associated to potential object detections. This speeds up computations and cuts the number of false-positive detections. The FOV is divided in $NxN$ cells with sizes 2.5 per cent of the FOV. We apply the refined quartile statistics described in Section 3.2 to select for each cell the best representative pixels of the local sky background, and compute the associated average $A$ and standard deviation $S$. For each cell, a local threshold $T = A + S$ is found. Then, the whole FOV is again divided – this time by much smaller $MxM$ pixel cells ($M=3$ pixels). Cells with even only 1 pixel below the local threshold $T$ are discarded. The surviving cells are associated to parts of bright or well-sampled objects, and to the central parts of faint ones. Many distinct cells with varying fluxes may be part of the same true object. Fig.1 displays a typical flux distribution of surviving cells.

From each surviving cell comes a detection of the entire object that contains that cell. Each detection is characterized by an aperture and sky background ring. The determination of optimal apertures and rings is described in Section 3.4. A single true object may have multiple detections at this stage. False-positive detections from leaking saturation flux and diffraction spikes, cosmic rays and other image artifacts may also be present. The selection of the best detection among multiple ones gives the final identification for the object. This and the elimination of false-positives are described in Section 3.5. Computation of object properties is given in Section 3.6.

### 3.4. Optimized apertures and sky background rings with BOIA

The best apertures and sky background rings are found by maximizing the S/N of the detection – BOIA's main concept. Starting from an initial 2-pixel radius, and centred at the $(x, y)$ coordinates of the input cell (Section 3.3), two 2-pixel-width circular rings of radii $r_1$ and $r_2 = r_1 + 2$ pixels are successively sampled, with their radius increased by 2 pixels at each iteration. F-tests and Student-t tests of significantly different means (Press et al., 1982) are applied to the pixel distributions of each ring, after eliminating higher/lower counts by quartile statistics (Section 3.2). Iterations stop when no significant difference between the average counts of the two adjacent rings is found, i.e. when the sky background around the object is reached. The last ring is the sky background ring (hereafter sb-ring). A new $(x, y)$ centre is computed by using a 2D version of the Modified Moment method (Stone (1989)) over the one-quarter brighter object pixels within the sb-ring inner radius. Starting from the new centre, the 2-rings procedure is repeated and we get a provisional inner radius $B$ for the sb-ring. The centre is again updated with the one-quarter brighter object pixels within $B$.

The object's aperture is delimited by $B$. The aperture radius $R_{ap}$ is sampled by $\Delta r = 0.1$ pixel steps from 1 up to $B$. The inner sb-ring radius $R_{sb}$ is sampled by 1 pixel steps from $B$ to $2B$, and its width $W$ by 1 pixel steps from 1 to 5 pixels. Following the photometric rationale by Howell (1989), the best $R_{ap}$, $R_{sb}$ and $W$ values give the highest S/N for the flux inside the aperture, subtracted from the sky background contribution from the pixels inside the sb-ring. Following Newberry (1991), the S/N is computed from Eq. 1, where $C$ (ADU units) is the clean flux of the object, $g$ is the gain, $n$ is the number of pixels associated to the object inside the aperture, $\sigma_{sb}$ is the sky background dispersion (ADU units) and $n_{sb}$ is the number of pixels used in the estimation of the sky background and dispersion after elimination of high/low counts by quartile statistics with sb-rings (Section 3.2).

$$\frac{S}{N} = \frac{\sqrt{C}}{\sqrt{\frac{1}{g} + n\,\sigma_{sb}^2\,C^{-1}\left(1 + \frac{1}{n_{sb}}\right)}} \quad (1)$$

We then recompute the centre, now using all the pixels inside the provisional optimal aperture radius $R_{ap}$ just found. Using this new and definite aperture centre and the same boundary value $B$ as before, we repeat the sampling to get the definite radius $R_{ap}$, sb-ring inner radius $R_{sb}$ and width $W$ of the detection.

### 3.5. Final object identification. Elimination of false-positives.

Definite identification comes by selecting one detection per object with elimination of remaining multiple detections and false-positives. This is done in the same order presented from Sections 3.5.1 to 3.5.3.

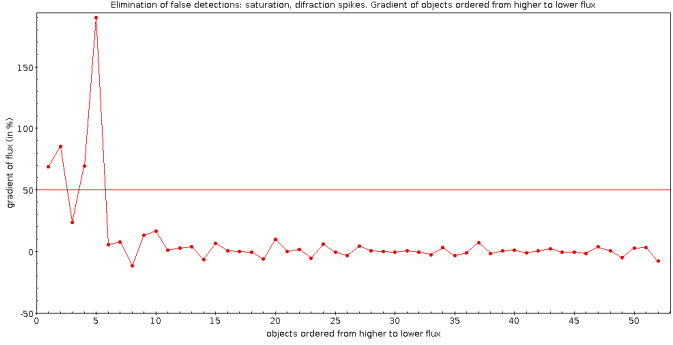
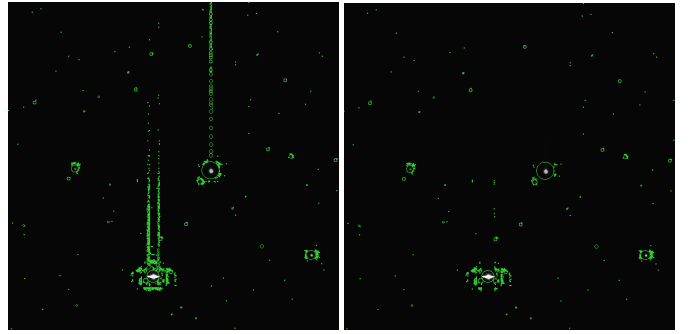

Figure 2: Flux gradient computed by Eq. 2 for a FOV with five bright sources that could generate false-positives by saturation leaking of diffraction spikes. Bright sources are searched for false-positive detections (see text).

Figure 3: Left panel: FOV with false-positives at the saturation leaking of bright sources. Right panel: same FOV after removal of the false-positives following the procedure described in the text.

*3.5.1. Elimination of multiple detections*

An expected outcome of object detection procedures (Section 3.3) is multiple detections per object. They are recognized by location: their centres lay inside the apertures of other multiples. For each object, only the detection with the largest aperture is preserved and associated parameters stored – the other multiple detections are discarded.

*3.5.2. Saturation leaking and diffraction spikes*

False-positives can be detected at the saturation leaking or diffraction spikes of bright sources. Their elimination starts by searching for bright sources. We order detections by flux $f$ and compute the gradient $G$ in accord to Eq. 2. From lower to higher fluxes, once a gradient above 50 per cent is found, source $i$ and all brighter ones are searched for the presence of false-positives. Fig.2 displays the computations for a FOV affected by the problem. In the example, the five brightest sources were selected for the searching of false-positives.

$$mean = \frac{f(i+1) + f(i-1)}{2} \; , \; G_i(\%) = 100 \, \frac{|f(i) - mean|}{f(i)} \quad (2)$$

Potential false-positives are vertically/horizontally aligned around the bright source. For vertical alignments, we determine the typical aperture size $R$ of detections found within 1 aperture radius from the $x$ bright source centre. We use quartile statistics with the elimination of the 25 per cent largest and smallest apertures to compute the average value $R$. We then sample these detections in bins of size $R$ along the $x$ axis, keeping the highest number $N$ of detections found in a bin. We do the same along the entire FOV, recording the number of vertically aligned detections for each bin $R$, and compute the average $n$ and standard deviation $sd$, but avoiding all the selected bright sources. The procedure is repeated for horizontal alignments, when the $y$ axis is probed instead. Vertical (horizontal) alignments are found if $N>5$ and the ratio $|N-n|/sd$ is higher than 10. In that case, all vertically (horizontally) aligned detections within 1 aperture radius from the $x$ ($y$) bright source centre are eliminated – the bright source is preserved. Fig.3 shows the same FOV used in the computations of Fig.2, before and after removal of false-positive detections. The procedure proved to be quite robust for empty and crowed star fields.

True-positives mistakenly eliminated, such as fainter Gaia stars or targets nearby bright sources, are later recovered by the procedure in Section 8.4, before the final $(x, y)$ to $(\alpha, \delta)$ reduction (Section 8.5).

*3.5.3. Faint detections, FOV crossers, cosmic rays, hot pixels*

Detections with apertures partially outside the FOV and faint spurious detections with the same radius for the aperture and inner sky background ring are eliminated at this stage.

Cosmic rays concentrate high fluxes in a small group of pixels, hot pixels in one. BOIA's original robust procedure for the elimination of cosmic rays and hot pixels is based on many empirical tests. We define contrast as the flux ratio with the squared aperture area (Eq. 3), not flux/area. We sort contrast in ascending order to get a contrast growth curve. Curves of true objects grow smoothly, while an abrupt enhance in contrast at the end signals cosmic rays and hot pixels.

$$Contrast = \frac{flux}{(aperture\ area)^2} \quad (3)$$

Starting from the middle of the contrast growth curve up, we apply the F-test in two sets: one with all growth curve points up to $i$ and another with $i$-1 points. If probability $p<0.05$, the variances of the sets are statistically different. Points $i$ and up are cosmic rays, marked and eliminated. Fig.4 shows an example.

*3.6. Object properties: final aperture centre, central moments, semi-major and minor axes, eccentricity and rotation angle*

BOIA's object astrometric/geometric properties – $(x, y)$ centre, semi-major axis $a$, semi-minor axis $b$ and orientation angle $\theta$ – are computed with the pixels inside BOIA's optimized apertures (Section 3.4). Given the object's image of pixel counts $I(x, y)$ and the smallest aperture count $C$, the reduced second order central moments $\mu'_{20}$, $\mu'_{11}$ and $\mu'_{02}$ of $I(x, y)$ are computed by the formulae in Eq. 4:

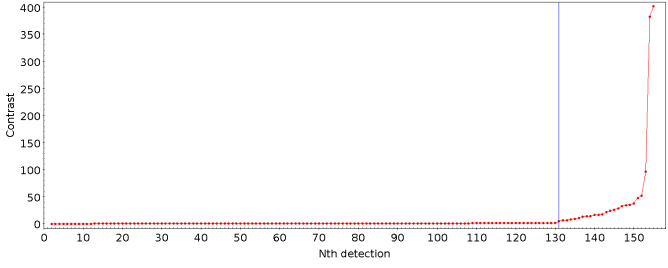

Figure 4: Contrast growth curve of detections as computed by Eq. 3. The F-test probability $p<0.05$ first occurs at point 131, i.e. the variances of all points up to 131 and up to 130 are statistically different. Points 131 and up are cosmic rays, a fact rigorously verified by visual inspection of the image.

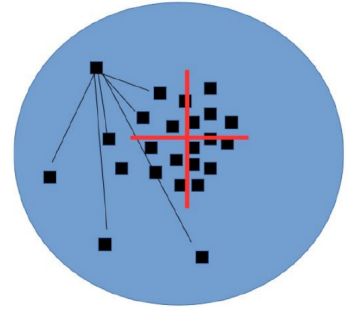

Figure 5: Illustration of a faint object observed under bad guiding. Some bright pixels above the threshold cluster around the "true" centre, while more isolated distant ones (indicated by lines) have lower weights $w(j,i)$. The red cross gives the Photogravity Center (PGM) – the natural choice "by eye". The Modified Moment (MM) centre (the blue circle centre) is clearly offset.

$$M_{ij} = \sum_x \sum_y x^i y^j \left[I(x,y) - C\right]$$

$$\mu'_{20} = \frac{M_{20}}{M_{00}} - \left(\frac{M_{10}}{M_{00}}\right)^2$$

$$\mu'_{11} = \frac{M_{11}}{M_{00}} - \left(\frac{M_{10}}{M_{00}}\right)\left(\frac{M_{01}}{M_{00}}\right)$$

$$\mu'_{02} = \frac{M_{02}}{M_{00}} - \left(\frac{M_{01}}{M_{00}}\right)^2 \qquad (4)$$

The semi-major axis $a$, semi-minor axis $b$, eccentricity $e$ and orientation angle $\theta$ (counterclockwise) of the semi-major axis $a$ with respect to the $x$ axis are computed from the reduced second order central moments with the formulae in Eq. 5 (for $\mu'_{20} = \mu'_{02}$ within $10^{-10}$ we have a circle and the indefinite angle $\theta$ is set to zero):

$$a = \frac{\mu'_{20} + \mu'_{02}}{2} + \frac{\sqrt{4\mu'^2_{11} + (\mu'_{20} - \mu'_{02})^2}}{2}$$

$$b = \frac{\mu'_{20} + \mu'_{02}}{2} - \frac{\sqrt{4\mu'^2_{11} + (\mu'_{20} - \mu'_{02})^2}}{2}$$

$$e = \sqrt{1 - \frac{b^2}{a^2}}, \quad \theta = \frac{1}{2}\arctan\left(\frac{2\mu'_{11}}{\mu'_{20} - \mu'_{02}}\right) \qquad (5)$$

The equivalent sigma $\sigma_E$ of the object is $\sigma_E = \sqrt{ab}$. For a perfect Normal profile, $\sigma_E$ tends to equal the Gaussian's statistical width $\sigma$ when most of the profile is contained inside the aperture. This frequently occurs for fainter objects – the most critical in PSF fitting. The centre is then further refined by using the novel Photogravity Center Method (PGM, see Section 4.1). The `BOIA` object properties are used for setting input values in the PSF fittings (see Section 4).

## 4. Measuring (x,y) centres

### 4.1. Photogravity Center Method (PGM)

The Modified Moment (MM, Stone 1989) weights $(x, y)$ pixel coordinates by counts above a threshold dependent of subjective sky background factors. The Photogravity Center Method (PGM) is an original novel algorithm developed by us that improves MM. Let $(j, i)$ be pixels associated to $(x, y)$ inside an aperture circle contained by a box of $(N_x, N_y)$ sizes, and $C(j, i)$ be ADU counts above a threshold $T$. Following quartile principles, the threshold $T$ is set from the 25 per cent brightest pixels. Besides ADU counts, PGM uses a weight $w(j, i)$ that privileges grouped pixels closer to each other, as in Eq. 6. The $(x_{PGM}, y_{PGM})$ measured centre is thus given by Eq. 7:

$$w(j,i) = \frac{1}{\sum_{m=1}^{N_X}\sum_{k=1}^{N_Y}[x(j,i)-x(m,k)]^2+[y(j,i)-y(m,k)]^2} \qquad (6)$$

$$x_{PGM} = \frac{\sum_{j=1}^{N_X}\sum_{i=1}^{N_Y} w(j,i)\, C(j,i)\, x(j,i)}{\sum_{k=1}^{N_X}\sum_{m=1}^{N_Y} w(k,m)\, C(k,m)}$$

$$y_{PGM} = \frac{\sum_{j=1}^{N_X}\sum_{i=1}^{N_Y} w(j,i)\, C(j,i)\, y(j,i)}{\sum_{k=1}^{N_X}\sum_{m=1}^{N_Y} w(k,m)\, C(k,m)} \qquad (7)$$

Hot pixels and telescope shifting by bad guiding result in large MM centroid offsets, mostly for faint objects. Fig. 5 illustrates an example. The "true" centre is the "gravity centre" around the clustered pixels, and is better "sensed" or "detected" by the PGM than with MM.

PGM $(x,y)$s, errors (Section 4.2) and object's properties (Section 3.6) are passed for $(\alpha, \delta)$ reductions and magnitude computations, otherwise they serve to set input values for PSF fittings (Section 4.3).

### 4.2. General (x,y) error formulae for Moment and PGM centres

We also compute realistic $(x, y)$ errors for PGM and MM centres – a feature made possible by `BOIA`, never seen before in the literature. By investigating the behaviour of $(x, y)$ errors from PSF fittings for a variety of magnitude ranges, telescopes, detectors, etc (see Section 10.1), we found a natural expression relating `BOIA`'s aperture parameters and image moments with the $(x, y)$ centre errors that we would expect for Moment centring methods. Such expression is given by the formulae in Eq. 8. From moment analysis computations (Section 3.6), $A$ and $B$ are normalized semi-major and semi-minor axes determined by the object's eccentricity $e$, $\theta$ is the orientation angle with respect to the $x$ axis, and $SN$ is the S/N of the aperture $R$.

$$A^2 = \frac{1}{\sqrt{1-e^2}} , \quad B^2 = \sqrt{1-e^2}$$
$$e_x = \sqrt{\pi} R \frac{\sqrt{A^2 cos^2\theta + B^2 sin^2\theta}}{SN}$$
$$e_y = \sqrt{\pi} R \frac{\sqrt{B^2 cos^2\theta + A^2 sin^2\theta}}{SN} \quad (8)$$

Eq. 8 expresses two basic ideas: moment centre errors should grow with the aperture area and get smaller as the S/N increases. For correctly dimensioning the error in pixel units (rather than in pixel$^2$ area units), the relation has the sense Error = $\sqrt{area}$ / SN, or Error = $\sqrt{\pi}$ R / SN for circular areas. Eq. 8 also takes into consideration the shape asymmetry of the object and how it propagates the error to the $x$ and $y$ directions. Here, $e = \sqrt{1-b^2/a^2} = \sqrt{1-B^2/A^2}$, $a$, $b$ being the semi-major and semi-minor axes and $e$ the eccentricity computed from the moment analysis in Section 3.6.

Based on aperture parameters computed from moment analysis, Eq. 8 is a valid general error formula for the $(x, y)$s obtained by PGM or any Moment centring methods, including the popular Modified Moment method (Stone, 1989), provided that we use the appropriate aperture radius, such as that from BOIA. This is a novel result, since non-model algorithms such as Moment methods are regarded as incapable of deriving $(x, y)$ error estimates (for a review of Moment methods see Zhang et al. 2021 and references therein). Notice that rather than derived analytically, Eq. 8 is checked successfully against experimental BOIA results, in a phenomenological sense.

### 4.3. Preparations for the PSF fittings

The sky background $S$ is taken from BOIA (not fitted). This recommended Least Squares (LS) procedure (Bevington, 1969) improves the sensitivity of the $\chi^2$ to the function parameters that actually shape the PSF model. For the LS kick-off (Press et al., 1982), the input values of some parameters common to all PSFs are computed in advance. From the PGM comes the $(x_0, y_0)$ centre. The height $h$ is set from the counts around the $(x_0, y_0)$ central pixel. A sanity check on $h$ avoids problems by saturated objects with unexpected low, negative or masked ADU counts. Assuming a Gaussian profile with height $H$ and statistical width equal to the equivalent sigma $\sigma_E$ (Section 3.6), and associating the volume $2\pi H \sigma_E^2$ to the total aperture flux $F$, we use the equivalent height $H = F/(2\pi\sigma_E^2)$ instead of $h$ if $H > h$.

### 4.4. Circular and Elliptical Gaussians

The 2D Circular Gaussian PSF model of the task $G_C(x, y)$ is given by Eq. 9, where $(x_0, y_0)$ is the centre, $h$ is the height above the flat sky background $S$ and $\sigma$ is the Gaussian statistical width. The input value of $\sigma$ for the LS kick-off comes from the equivalent sigma $\sigma_E$ (Section 3.6).

$$G_C(x, y) = h.e^{-\left(\frac{(x-x_0)^2+(y-y_0)^2}{2.\sigma^2}\right)} + S \quad (9)$$

The 2D elliptical Gaussian PSF model of the task $G_E(x, y)$ is given by Eq. 10, where $(x_0, y_0)$ is the centre, $h$ is the height above the flat sky background $S$, and $A$, $B$ and $C$ are coefficients related to the rotation angle $\theta$ and to the semi-major and semi-minor axes which define the elliptical profile widths. The input values of $A$, $B$ and $C$ for the LS kick-off are calculated with the formulae in Eq. 11, where the semi-major axis $a$, semi-minor axis $b$ and the rotation angle $\theta$ of the semi-major axis $a$ with respect to the $x$ axis come from the central moment analysis (Section 3.6).

$$G_E(x, y) = h.e^{-\left(\frac{A(x-x_0)^2+B(x-x_0)(y-y_0)+C(y-y_0)^2}{2}\right)} + S \quad (10)$$

$$A = \left(\frac{\cos\theta}{a}\right)^2 + \left(\frac{\sin\theta}{b}\right)^2$$
$$B = -2\cos\theta\sin\theta\left(\frac{1}{b^2} - \frac{1}{a^2}\right)^2$$
$$C = \left(\frac{\cos\theta}{b}\right)^2 + \left(\frac{\sin\theta}{a}\right) \quad (11)$$

Fitted $a$, $b$ and $\theta$ values come from the fitted $A$, $B$ and $C$ coefficients in accord to the formulae in Eq. 12.

$$a = -\frac{\sqrt{2(B^2-4AC)([A+C]+\sqrt{(A-C)^2+B^2})}}{B^2-4AC}$$
$$b = -\frac{\sqrt{2(B^2-4AC)([A+C]-\sqrt{(A-C)^2+B^2})}}{B^2-4AC}$$
$$\theta = \begin{cases} \arctan\left(\frac{(C-A-\sqrt{(A-C)^2+B^2})}{B}\right) & for\ B \neq 0 \\ 0 & for\ B = 0, A < C \\ 90° & for\ B = 0, A > C \end{cases} \quad (12)$$

### 4.5. Circular and Elliptical Lorentzians

The 2D Circular Lorentzian PSF model $L_C(x, y)$ is given by Eq. 13, where $(x_0, y_0)$ is the centre, $h$ is the height above the flat sky background $S$, $r_0$ is the Lorentzian width, and $\alpha$ and $\beta$ are coefficients that shape respectively the centre and wings of the PSF. LS kick-off input values for the Lorentzian width $r_0$ and the $\alpha$ and $\beta$ coefficients are computed simultaneously. Since the Gaussian PSF is a good approximation for well-behaved stellar profiles, and since Lorentzian PSFs are similar to the Gaussian ones at the centre for $\alpha=2$, we start with this $\alpha$. From empirical testing, we also start with $\beta=1$ for the same reasons. The procedure samples pixel rings from the centre toward the aperture borders until the counts reach $h(2^\beta)^{-1}$ (i.e. $h/2$ in the first iteration) when $r_0$ is then found. The value of $\alpha$ is then reset for the value found for $r_0$. The process can be iterated, but tests showed that further iterations are not needed besides those inherent to the LS. Thus, for saving computer time, the values found at this stage for $r_0$ and $\alpha$ with $\beta = 1$ are used as kick-off for the LS. After the LS fitting, the Lorentzian width $r_0$ is scaled by the factor $(2\beta \ln 2)^{-\frac{1}{2}}$ for comparison with the Gaussian statistical width $\sigma$.

$$L_C(x,y) = \frac{h}{\left(\frac{[(x-x_0)^2+(y-y_0)^2]^{\frac{\alpha}{2}}}{r_0^\alpha} + 1\right)^\beta} + S \qquad (13)$$

The 2D elliptical Lorentzian PSF model $L_E(x,y)$ is given by Eq. 14, where $(x_0, y_0)$ is the centre, $h$ is the height above the flat sky background $S$, $\alpha$ and $\beta$ are the characteristic coefficients of the Lorentzian profile, and $A$, $B$ and $C$ are coefficients related to the rotation angle $\theta$, the semi-major and semi-minor axes which define the elliptical profile widths. LS kick-off input values for $\alpha$ and $\beta$ come from a preliminary circular Lorentzian PSF fitting. Input values for $A$, $B$ and $C$ are computed from $a$, $b$ and $\theta$ input values by using the same formulae of Eq. 11 (Section 4.4); $a$, $b$ and $\theta$ come from the central moment analysis (Section 3.6), but $a$ and $b$ are first scaled by $r_0/\sigma_E$, with $r_0$ coming from the preliminary circular Lorentzian fit and $\sigma_E$ being the equivalent width from the central moment analysis.

$$L_E(x,y) = \frac{h}{\left([A(x-x_0)^2+B(x-x_0)(y-y_0)+C(y-y_0)^2]^{\frac{\alpha}{2}}+1\right)^\beta} + S \qquad (14)$$

Fitted $a$, $b$ and $\theta$ values come from the fitted $A$, $B$ and $C$ coefficients by using the same formulae of Eq. 12 (Section 4.4). For comparison with elliptical Gaussian statistical width semi-major and semi-minor axes, the Lorentzian semi-major and semi-minor axes $a$ and $b$ are also scaled by the factor $(2\beta \ln 2)^{-\frac{1}{2}}$.

By setting $\alpha = 2$, Eq. 13 reduces to the Moffat profile (Moffat, 1969). Fig. 6 compares elliptical Lorentzian and Gaussian profiles with $\alpha = 2$, $\beta = 1$, equal heights, rotation angles, eccentricity and semi-axes widths at the half maximum height. We get equal Lorentzian and Gaussian profiles at the centre, but Lorentzian wings are more pronounced and better resemble stellar images in digitized plates and CCD observations. The generalized Lorentzian PSFs of Eq. 13 and Eq. 14 with the $\alpha$ coefficient give even more flexibility than the Moffat PSF, since we can shape the central parts too. The roles of the $\alpha$ and $\beta$ coefficients of these PSFs are depicted in Fig. 7. Generalized Lorentzian PSFs not only can model a variety of wing shapes, but also fit distinct centre profiles, including non-linear regimes and even saturated images.

### 4.6. Floor PSF, elimination of multiple (x,y)s, small FWHMs

Prior to the $(\alpha, \delta)$ reductions (Section 8), we must eliminate spurious objects that may have eventually survived BOIA procedures.

The Floor PSF serves for identifying spurious detections submitted for PSF fitting. It is a bivariate $(x, y)$ plane. The fitting errors of the Floor and $(x, y)$ PSFs are compared by a F-test (Press et al., 1982). If probability $p > 0.6171$, i.e. if errors are close in a $0.5\,\sigma$ sense, both PSF fittings are equivalent. No object exists inside the aperture or it is too faint for a successful PSF fit, and the measurement is discarded.

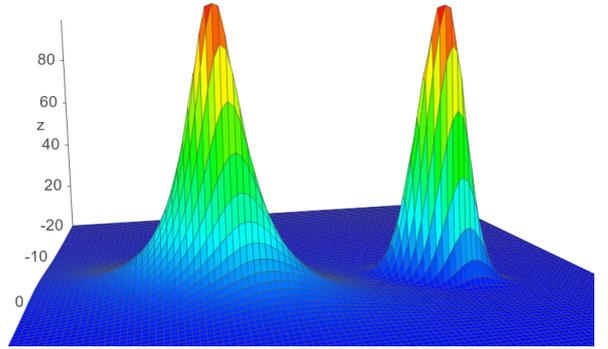

Figure 6: Elliptical Lorentzian (left) and Gaussian (right) profiles in accord to Eq. 14 and Eq. 10 respectively with the same height, rotation angle, eccentricity ($e$=0.2) and semi-axes widths at half maximum height. Here $\alpha$=2 and $\beta$=1 for the Lorentzian PSF. Notice the more pronounced wings of the Lorentzian profile, better resembling actual stellar images from digitized plates and CCD observations. Similar plots are obtained for circular profiles.

Sometimes close detections of the same object pass through the elimination filters in the BOIA procedures. After the definite $(x, y)$ measurements, the centres of the same object tend to converge and finally intersect each other within the apertures. For eliminating multiple $(x, y)$ measurements of the same object, the same procedure described in Section 3.5.1 for the elimination of multiple detections by centre intersection is repeated. Only the larger aperture survives.

We also eliminate spurious objects with small discrepant FWHMs. We sort objects by *seeing*, apply quartile statistics and set a $5\sigma$ threshold for the smaller FWHMs, below which any object is discarded.

## 5. Catalogues

PAT works with up to three reference catalogues: the Gaia DR3 (Gaia Collaboration, 2021), the User Catalogue and the Ephemeris Catalogue. 2MASS (Cutri et al., 2003) gives $J$, $H$ and $K$ star magnitudes.

Gaia DR3 is the main PAT reference catalogue. It is the standard materialization of the ICRS system in the optical domain. Most stars have five of the six astrometric parameters (the sixth being the radial velocity) needed for converting ICRS coordinates to astrometric or apparent ones, with accuracy/precision better than 1 mas for $8 < G$ magnitude $< 21$, with uniform star coverage and completeness. Due to small rotation frame biases, bright $G < 13$ star proper motions are corrected to the quasar's standard rest frame following Cantat-Gaudin & Brandt (2021). PAT can drop Gaia stars by magnitude range, absence of proper motions, parallax, radial velocity or duplicity flag. PAT will absorb future Gaia releases and astrometric parameter upgrades.

Another PAT reference catalogue is the User Catalogue of positions and magnitudes set by the user for any type of object (stars, quasars, galaxies). Proper motions, parallax and radial velocities are optional.

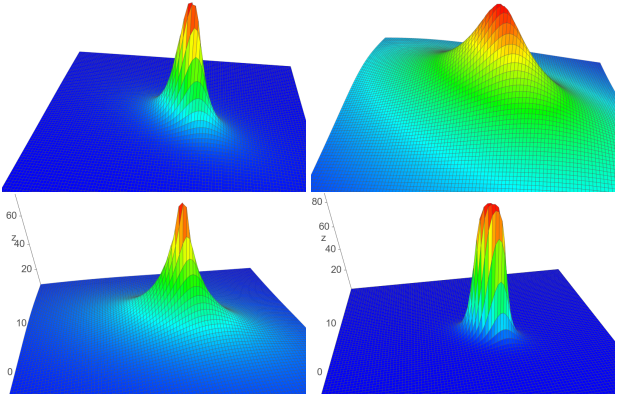

Figure 7: Elliptical Lorentzian profiles in accord to Eq. 14. The $\alpha$ coefficient shapes the centre while $\beta$ rules the wings. The same width at half maximum height is set for all plots. Top left panel has $\alpha = 2$ and $\beta=1$ which seems to better "mimic" the Gaussian profile. Top right panel has $\alpha = 2$ and $\beta=0.1$ showing that the wings are raised by smaller $\beta$ values and vice-versa. Lower left panel has $\beta=1$ and $\alpha=1$, and lower right has $\beta=1$ and $\alpha=4$, showing that for $\alpha<2$ we get somewhat unrealistic spike-shaped profiles at the centre and for $\alpha>2$ we can easily model non-linear regimes and even saturated profiles. Similar plots are obtained for circular profiles.

The Ephemeris Catalogue is suited for the astrometry of natural satellites in the absence of reference stars. The reference frame is given by the satellite positions. Useful in the astrometry of the Saturn and Uranus systems, the catalogue has magnitudes and $(\alpha, \delta)$ positions for the instants of the observations.

Gaia and 2MASS data are automatically accessed from the Vizier database at CDS (Ochsenbein, Bauer, & Marcout, 2000) using the CDSCLIENT tool[6]. The default size of the extracted FOV is $2^o$x$2^o$, changeable. Catalogue data files are temporarily stored in the hard disk, then erased after execution. During the task run, for FOVs centred within 1 arc minute from each other, no new internet access is made and the downloaded data of the first FOV is reused for gaining processing speed. The User and Ephemeris catalogue files must be available in the local hard disk. They must serve for all FOVs treated for each task run. An auxiliary PRAIA task creates PAT Ephemeris catalogues (see Section 12). FOV automatic reference object identification uses one of the available catalogues in hierarchical order: Gaia, User and Ephemeris ones (Section 8.2).

## 6. Targets

PAT targets may have fixed or varying coordinates. Objects with fixed coordinates may be stars, quasars, galaxies, etc. Their set coordinates are valid for any observation date. No proper motions, parallax, etc are given – they are absorbed in the input position. Each entry in the target input list will be searched for all FOVs. Objects with varying coordinates are interpreted as Solar System ones. Their individual positions are only valid for each FOV with an specific mid-instant of observation. Other information like estimated magnitude, solar phase angle and wavelength of each observation are also necessary. The same auxiliary PRAIA task for ephemeris extraction is used for preparing the input target list of Solar System bodies (Section 12). The identification of targets in the FOV is made straightforwardly by the $(\alpha, \delta)$ and instant of measured positions, with each fixed target searched for in all images. The error cube $(\Delta\alpha, \Delta\delta, \Delta t)$ for target identification is set by the user.

## 7. Catalogue and target positions to date

The reference catalogue positions used in the tangent plane polynomial fittings (Section 8) are apparent places at the date of observation affected by atmospheric refraction, allowing for better fitting/interpretation of the computed model coefficients.

Gaia and User Catalogue ICRS positions are reduced to apparent places at observation date and affected by refraction following IAU rules with SOFA routines (IAU SOFA Board, 2021).

The Ephemeris Catalogue has ICRS astrometric topocentric coordinates reduced to airless apparent places at observation date affected by light time delay, gravitational deflection of light by the Sun, Jupiter and Saturn, precession and nutation, and stellar aberration in accord to the models and definitions used by the Jet Propulsion Laboratory (JPL) – see the JPL Web Horizons Service[7]. The extracted airless apparent ephemeris is further affected by solar phase angle following Lindegren (1977), and by refraction (IAU SOFA Board, 2021).

All input target ICRS astrometric topocentric positions are reduced to airless apparent places at the date of observation, and affected by atmospheric refraction. For fixed objects, it is done with SOFA routines (IAU SOFA Board, 2021). For moving Solar System bodies, all operations regarding to the Ephemeris Catalogue also apply.

## 8. Reduction to $(\alpha,\delta)$

After object detection (Section 3) and $(x, y)$ measurements (Section 4), we identify reference catalogue stars and make $(x, y)$ to $(\alpha, \delta)$ reductions. Identification is fully automatic, dispensing any information, even the pixel scale (Section 8.2). The $(x, y)$ to $(\alpha, \delta)$ reduction is described in Section 8.3. It is made two times with a whole new round of $(x, y)$ measurements made in between for further improving the centres and recovering missing objects.

After a primary $(x, y)$ to $(\alpha, \delta)$ reduction (Section 8.3), we invert the sense and reduce $(\alpha, \delta)$s to $(x, y)$s for finding the refined location of all previously measured objects. This also allows for the search of missing catalogue reference objects and targets with no previous measured counterparts on their expected $(x, y)$ coordinates. We then compute new (possibly) improved BOIA-like aperture parameters for all objects (missing or not), based on the previous BOIA measurements of reference objects. Then, a new round of $(x, y)$ measurements is done, including already detected objects and missing ones in an attempt to recover them. Among the old and new $(x, y)$s of the former detected objects, the ones with best centre errors are preserved. The recovered objects have their $(x, y)$ measurements and related data recorded. The procedure is detailed in Section 8.4. A final $(x, y)$ to $(\alpha, \delta)$ reduction (Section 8.5) is then performed using the final $(x, y)$ measurements of all previously detected and recovered objects.

---

[6]Vizier CDSCLIENT: http://cdsweb.u-strasbg.fr/doc/cdsclient.html

[7]JPL Horizons Web Ephemeris Service: https://ssd.jpl.nasa.gov/horizons/app.html#/

PAT also reduces positions with $(x, y)$ and $(\alpha, \delta)$ inputs (no images) from other PAT runs or other packages (Section 8.6).

*8.1. Plane/spherical position conversions; polynomial models*

Conversion between spherical $(\alpha, \delta)$ and linear standard tangent plane $(X, Y)$ coordinates follows Gnomonic (Eq. 15) and Anti-Gnomonic (Eq. 16) projections, $(\alpha_0, \delta_0)$ being the tangent point coordinates. Table 1 lists the polynomial models available by PAT relating $(X, Y)$ coordinates with $(x, y)$ measurements.

$$X = \frac{cos\delta \, sin(\alpha - \alpha_0)}{sin\delta \, sin\delta_0 + cos\delta \, cos\delta_0 \, cos(\alpha - \alpha_0)}$$
$$Y = \frac{sin\delta \, cos\delta_0 - cos\delta \, sin\delta_0 \, cos(\alpha - \alpha_0)}{sin\delta \, sin\delta_0 + cos\delta \, cos\delta_0 \, cos(\alpha - \alpha_0)} \quad (15)$$

$$tan(\alpha - \alpha_0) = \frac{X}{cos\delta_0 - Y sin\delta_0}$$
$$tan\,\delta = \frac{(Y cos\delta_0 + sin\delta_0) \, cos(\alpha - \alpha_0)}{cos\delta_0 - Y sin\delta_0} \quad (16)$$

*8.2. Automatic identification of reference catalogue objects*

The identification of catalogue reference objects among the measurements is needed for mapping transformations between $(x, y)$ and $(\alpha, \delta)$ coordinates. In previous task versions, the pixel scale and error were needed a priori. Bright measured and catalog object pairs were sampled and the distance ratios in arcseconds per pixel were checked against the pixel scale, to find matches within the error margin. However, sometimes the pixel scale, or its error, or both were not exactly known, or not known at all. Sometimes the procedure simply failed, mostly in crowded star fields, demanding too much intervention to setup new measured/catalogue bright object proportions.

Pixel scale, error and any other information are no longer needed. The $(x, y)$ to $(\alpha, \delta)$ mapping is now fully done in a fast and robust automatic fashion. We use triangles and test matches on 5 distinct sky scales with the help of a voting scheme. After the identification of the reference catalogue objects, the pixel scale and its error are deduced, among other useful FOV information like sizes and orientation angle.

The identification procedure is as follows. We select the *NM* brightest measured objects and *NC* brightest catalogue objects with *NM* < *NC*. We perform all possible *permutations* of measured objects and *combinations* of catalogue objects to form triangles, computing and storing the angles of each vertex *A*, *B* and *C* as in Eq. 17, where *a*, *b* and *c* are their associated triangle sides as in Fig.8. The triangle sides of measured objects are computed from their $(x, y)$ coordinates. Those of catalogue

Table 1: PRAIA astrometry task polynomial models that relate tangent plane $(X, Y)$ coordinates with $(x, y)$ measurements.

| Model | X | Y |
|---|---|---|
| M1 | $a_{00} + a_{10}\,x + a_{01}\,y$ | $b_{00} - a_{01}\,x + a_{10}\,y$ |
| M2 | $\sum_{j+i=0}^{1} a_{ji}\,x^j y^i$ | $\sum_{j+i=0}^{1} b_{ji}\,x^j y^i$ |
| M3 | $\sum_{j+i=0}^{2} a_{ji}\,x^j y^i$ | $\sum_{j+i=0}^{2} b_{ji}\,x^j y^i$ |
| M4 | M3(X) $+a_3\,x(x^2+y^2)$ | M3(Y) $+b_3\,y(x^2+y^2)$ |
| M5 | M4(X) $+a_5\,x(x^2+y^2)^2$ | M4(Y) $+b_5\,y(x^2+y^2)^2$ |
| M6 | $\sum_{j+i=0}^{3} a_{ji}\,x^j y^i$ | $\sum_{j+i=0}^{3} b_{ji}\,x^j y^i$ |
| M7 | M6(X) $+a_5\,x(x^2+y^2)^2$ | M6(Y) $+b_5\,y(x^2+y^2)^2$ |
| M8 | $\sum_{j+i=0}^{5} a_{ji}\,x^j y^i$ | $\sum_{j+i=0}^{5} b_{ji}\,x^j y^i$ |

*Notes*: M1 = Four Constants model; M2 = Complete First Degree model; M3 = Complete Second Degree model; M4 = M3 model + radial distortion terms of third order; M5 = M4 model + radial distortion terms of fifth order; M6 = Complete Third Degree model; M7 = M6 model + radial distortion terms of fifth order; M8 = Complete Fifth Degree model. Models are numbered in hierarchical order with respect to the identification of reference catalogue objects (see Section 8.2).

objects come from their standard $(X, Y)$ coordinates, obtained by the Gnomonic projection of $(\alpha, \delta)$ coordinates (Eq. 15).

$$A = \arccos\left(\frac{b^2+c^2-a^2}{2\,b\,c}\right), B = \arccos\left(\frac{a^2+c^2-b^2}{2\,a\,c}\right), C = \arccos\left(\frac{a^2+b^2-c^2}{2\,a\,b}\right) \quad (17)$$

A measure/catalogue triangle match occurs whenever the respective vertex angles *A*, *B* and *C* of each measured and catalogue triangle differ by less than 1 arc minute. We test all matched measure/catalogue triangles with Four Constants polynomial model fittings in the tangent plane (Table 1). Since FOV sizes are not known a priori, the procedure is repeated 5 times for hierarchically half smaller FOVs, starting with $2^o$x$2^o$, then $1^o$x$1^o$, 30'x30', 15'x15' and 7.5'x7.5'. But we can setup a different initial FOV size. From all 5 sampled FOV sizes, the matching triangle pair with more recognized common measured/catalogue bright objects is selected.

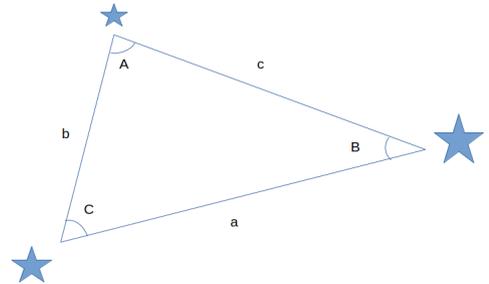

Figure 8: Nomenclature convention for the vertex angles and sides of tangent plane triangles for the calculation of angles by Eq.17 in the identification of common catalogue/measured objects (see text).

Next, using all newly identified common measured/catalogue bright objects (besides the ones of the triangle pair), we make a new fit with the complete First Degree polynomial model (Table 1), compute $(X, Y)$ standard coordinates for all measured objects in the FOV and search all catalogue objects for new identifications. We repeat the fittings, incorporating new common objects and incrementing the tangent plane model in hierarchical order (see Table 1). We stop after reaching the set model.

Experience shows that 15 brightest measured objects and 60 catalogue ones suffice for the identification of FOVs varying from 3'x3' up to $2^o$x$2^o$ in an extremely fast way, taking only 1 – 2 s in a i7-core computer for completing the 5 cycles.

After identifying all common measured/catalogue objects, we compute and display provisional values for the pixel scale and its error, FOV sizes and the orientation angle of the $x$ axis with respect to the right ascension direction.

### 8.3. Primary reduction of $(x, y)$ to $(\alpha, \delta)$

Using the catalogue reference objects measured in the FOV, we compute standard coordinates $(X, Y)$ in the tangent plane from $(\alpha, \delta)$ coordinates by using the Gnomonic projection (Eq. 15). We then use the chosen polynomial model (Table 1) for fitting the $(x, y)$ measurements of reference objects to their $(X, Y)$ coordinates by a Levenberg–Marquardt Least-Squares procedure (Press et al., 1982). Outliers are eliminated one by one. PAT can stop elimination when the highest |O-C| is smaller than a given threshold, or by sigma-clipping. Using the fitted polynomials, all measured $(x, y)$s are reduced to $(X, Y)$ coordinates, which are converted to $(\alpha, \delta)$ by the Anti-Gnomonic projection (Eq. 16). In this step, the $(x, y)$ centre is also added and subtracted by the $(e_x, e_y)$ measurement errors in pixels. Computed auxiliary $(\alpha, \delta)$s are subtracted from the object's $(\alpha, \delta)$, and the largest difference in each coordinate is kept as the measured $(\alpha, \delta)$ error in mas. This provides more accurate $(\alpha, \delta)$ measurement error values than multiplying the pixel scale by $(e_x, e_y)$, as it does take into account local distortions in the FOV. Standard errors $(E_\alpha, E_\delta)$ are also computed from the polynomial Least-Squares fits (Eichhorn & Williams, 1963). The pixel scale, error, FOV sizes and orientation angle of the $x$ axis with respect to the $\alpha$ direction are also computed and displayed.

### 8.4. Re-measuring and recovering all catalogue objects/targets

After the primary $(x, y)$ to $(\alpha, \delta)$ reduction (Section 8.3), we attempt to refine the $(x, y)$ measurements of all catalogue reference objects and targets before a second and definite $(\alpha, \delta)$ reduction. Based on the BOIA aperture sizes found for catalogue reference objects, we can relate aperture dimensions and magnitudes in a robust way. We can also refine the input aperture $(x, y)$ centres for catalogue and target objects by inverting the sense of the reduction from $(\alpha, \delta)$ to $(x, y)$. Using refined aperture sizes and input $(x, y)$ centres, we try to improve the $(x, y)$ measurements of catalogue and target objects.

Inherent to the procedure is the recovering of missing reference catalogue objects and targets. An object is missing if no $(x, y)$ counterpart is found among all measured objects in the FOV within a given error box, after the inversion from $(\alpha, \delta)$ to $(x, y)$ coordinates. Using extrapolated BOIA aperture sizes and the inverted $(x, y)$ coordinates of missing objects, we try to measure and recover them for the final $(\alpha, \delta)$ reduction.

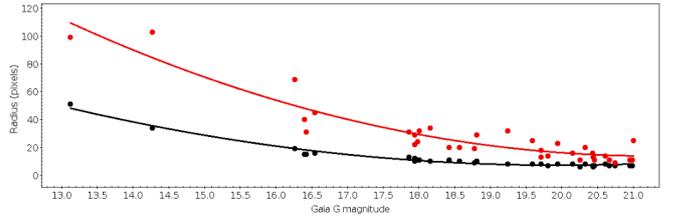

Figure 9: Fittings for the BOIA aperture radius and inner sky background ring radius by the magnitude empirical relation of Eq. 18 with Gaia stars. Black refers to the aperture radius and red to the inner sky background ring radius. From known magnitudes, the BOIA-like optimum aperture radius and inner sky background ring radius of any object can be estimated.

Taking catalogue reference objects identified in the FOV and used in the primary reduction, we compute $(X, Y)$ coordinates from $(\alpha, \delta)$ by the Gnomonic projection (Eq. 15). We then use the same tangent plane model but invert the sense of the fitting to find the relation that reduces $(X, Y)$ coordinates to $(x, y)$ measured ones, i.e. we take the formal polynomial expressions in Table 1, exchanging $(x, y)$ by $(X, Y)$ and vice-versa. This accurately pins down the $(x, y)$ location of any object in the FOV from their known $(\alpha, \delta)$ coordinates.

The empirical function in Eq. 18 relates the aperture radius with magnitudes, where $A, B, C$ are linear coefficients. By using the magnitude of reference catalogue objects and their BOIA optimum aperture radii (Section 3.4), we fit Eq. 18 to get the BOIA-like optimum aperture radius from the known magnitude of any object, including missing ones. The same relation serves to find the inner radius of sky background rings. The sky background ring width is fixed in 2 pixels (an usual value found by BOIA). We also limit the radius of the object's apertures and inner sky background rings to the fitted values for the brighter and fainter used catalogue reference objects. This avoids divergence problems in the fitting outside the magnitude range of the used reference objects. Fig.9 displays an example using Gaia stars. Since the apertures and sky background rings used in the fittings are optimal in a BOIA sense (Section 3.4), so are apertures and sky background rings estimated by this procedure.

$$Radius = A + B\ mag + C\ mag^2 \qquad (18)$$

Using the optimal apertures/rings and inverted $(x, y)$ coordinates found for all reference catalogue objects and targets, we proceed to a new determination of object image properties (Section 3.6) and perform new $(x, y)$ measurements (Section 4). Missing objects are then recovered with their $(x, y)$ centres and properties measured for the first time. For objects with new and old $(x, y)$ measurements, the ones with the smallest $(x, y)$ error are preserved with the related object properties – the other coordinates and aperture data are discarded.

## 8.5. Final reduction of (x, y) to (α, δ)

After remeasuring reference catalogue objects and targets and recovering missing ones (Section 8.4), a new $(x, y)$ to $(\alpha, \delta)$ reduction is done for the last time with all updated $(x, y)$ centres, including recovered catalogue reference stars and targets. The procedure is identical as that described in Section 8.3 for the primary $(x, y)$ to $(\alpha, \delta)$ reduction. Final values for the pixel scale, error, FOV sizes and orientation angle of $x$ with respect to the right ascension direction are computed and displayed.

## 8.6. Reduction from previous (x, y) or (α, δ) measurements

PAT also determines positions from $(x, y)$ or $(\alpha, \delta)$ measurements (not images) from previous PAT runs or other packages. This is useful to remake astrometry made with old/other reference catalogues, or to quickly test different PAT parameters in the astrometry of previously measured FOVs without needing to repeat the image processing. Reference catalogue objects are directly identified from the available input $(\alpha, \delta)$s, with minimum/maximum mid-values furnishing tangent plane coordinates. A unique reduction furnishes all the final $(\alpha, \delta)$ positions from the input $(x, y)$s, following the procedures described in Section 8.3. If only $(\alpha, \delta)$ inputs are available, PAT uses the corresponding $(X, Y)$ tangent plane standard coordinates computed by Gnomic projection (Eq. 15) in place of $(x, y)$s. PAT also accepts field distortion pattern inputs created externally or by previous runs with the task to improve astrometry (see the User Manual).

## 9. Positions, magnitudes, errors, visualisation

All output positions are converted from apparent to astrometric topocentric coordinates, including reference catalogue object (O-C)s and ephemeris position offsets. For non Solar System objects, conversion includes corrections by refraction and follow IAU recommendations through the SOFA routines (IAU SOFA Board, 2021). For Solar System bodies, besides refraction, it includes corrections by solar phase angle (Lindegren, 1977), stellar aberration, precession and nutation, gravitational deflection of light by the Sun, Jupiter and Saturn, and light time delay, following JPL's definitions.

Output magnitudes $m$ are given by $m=k-2.5\log_{10}(F)$. The zero point $k$ is computed with the magnitude of reference catalogue objects – G band for Gaia DR3 stars – and with the total flux $F$ from PSF fittings or BOIA aperture photometry. Outliers are eliminated by sigma-clipping. Mean errors come from the standard deviation of magnitude (O-C)s.

Astrometric data are output for all measured objects: $(\alpha, \delta)$ positions, standard errors $(E_\alpha, E_\delta)$, measurement errors $(e_\alpha, e_\delta)$, mean errors $(\sigma_\alpha, \sigma_\delta)$, (O-C)s of reference catalogue objects (including flags for outliers), FWHMs, catalogue and measured magnitudes and errors (including infrared 2MASS magnitudes), instants. No total error for the $(\alpha, \delta)$ positions is furnished – this is a controversial topic – but all kinds of error estimates and related values are available, so that one can compute total errors in accord to one's own judgment.

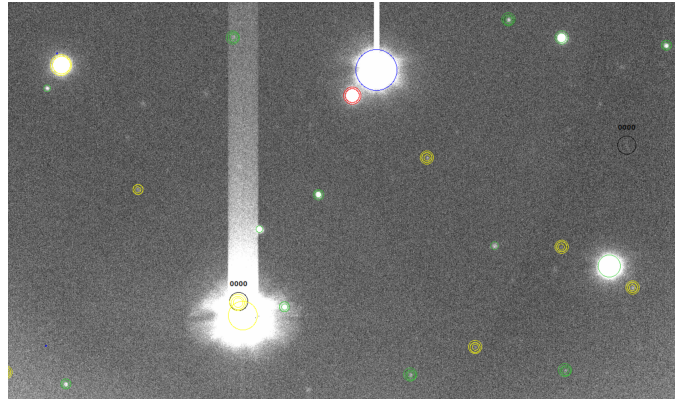

Figure 10: FOV with Triton, Neptune's main satellite (see FOV 1 in Section 10.1). Circles are centred at the measured $(x, y)$s with aperture sizes as determined by BOIA procedures. Targets (Triton here) are indicated by red circles. Green and yellow colors respectively indicate used and discarded (high |O-C|s) Gaia reference stars. Blue circles (e.g. Neptune) regard to measurements that are neither targets nor reference catalogue objects. Gaia stars not successfully identified/measured are indicated by black circles with numeric codes. Single, double and triple circles regard to the $(\alpha, \delta)$ origin (BOIA measurement or recovery, primary or final reduction position).

Target "measured - reference" $(\alpha, \delta)$ offsets are given, and separate files output in PRAIA, MPC and NIMA (Numerical Integration of the Motion of an Asteroid, Desmars et al. 2015) formats. For visual inspection analysis, PAT outputs files in ds9 region format. Colored circles indicate object type, centring and $(\alpha, \delta)$ reduction information, as explained in Fig. 10.

## 10. PAT validation: position error analysis

Here we illustrate the astrometric performance of the current PAT version through a detailed analysis of measurement and position errors. The analysis is based on the astrometric treatment of typical and often challenging observations that represent the variety of cases in routine work by the Rio Group and the Lucky Star collaboration.

### 10.1. FOV samples for evaluating (x,y) errors and object widths

For illustrating the results of exhaustive tests over a variety of instruments, magnitude ranges and star distributions in the sky, we present 5 FOVs that display typical and sometimes challenging astrometric and photometric characteristics present in most CCD observations. The studies presented next in Sections 10.2 and 10.3 about the object widths and measured $(x, y)$ errors with PAT refer to these FOVs, displayed and described in detail in Fig.11, with main characteristics and $(x, y)$ to $(\alpha, \delta)$ reduction information given in Table 2 and Table 3.

### 10.2. PGM/PSF object widths and the BOIA aperture radius

The analysis of the PGM and PSF object widths and their relation with the BOIA aperture radius is instructive. Taking Gaia stars with known magnitudes, Fig.12 shows the variation of the FWHM with brightness for the PGM and each PSF (Circular and Elliptical Gaussian and Lorentzian models), for each of the 5 FOVs described in Section 10.1. For elliptical PSFs,

Table 2: Sampled FOVs covering a variety of instruments, magnitudes and star distributions in the sky (see FOV details in Fig.11).

| FOV | $(\alpha, \delta)$ | Sizes | Scale | Gaia | CCD | Telescope | CCD | Filter |
|---|---|---|---|---|---|---|---|---|
| | (h,dg) | (min) | "/pixel | N/min$^2$ | gain | IAU code, D | pixels, model | |
| FOV 1 | 06 +24 | 2.9 x 2.9 | 0.170 | 16.4 | 19.3 | (874) OPD T1.6m | 1024$^2$ Ixon Em | I |
| FOV 2 | 23 −08 | 6.0 x 6.0 | 0.177 | 1.1 | 4.0 | (874) OPD T1.6m | 2048$^2$ Ikon | I |
| FOV 3 | 03 −55 | $\pi 7.2^2$ | 0.290 | 0.3 | 2.67 | (I33) SOAR T4m | 1548$^2$ Goodman | iSDSS |
| FOV 4 | 18 −19 | 12.0 x 12.0 | 0.350 | 101.1 | 3.5 | (874) OPD T0.6m | 2048$^2$ Ikon | I |
| FOV 5 | 07 +12 | 73.7 x 73.7 | 1.079 | 9.0/5.5* | 1.49 | (71) Rozhen T0.5m | 4096$^2$ FLI | R |

*Notes*: The FOV of the SOAR Goodman Imager (a Fairchild CCD) was binned (2x2) and is actually a circle with R = 7.2'. FOV 5 was observed at the Rozhen National Astronomical Observatory of Bulgaria with a Finger Lakes Instrumentation ProLine CCD camera ON Semi KAF-16803 model. (*) FOV 5 has 48999 Gaia stars (9.0 stars per arcminute$^2$), but only G < 19.5 stars (5.5 stars per arcminute$^2$) were used in the $(x, y)$ to $(\alpha, \delta)$ reductions (see Table 3).

Table 3: Information about the $(x, y)$ to $(\alpha, \delta)$ reduction of the sampled FOVs (see FOV details in Fig.11).

| | | | PGM | | | CGA | | | EGA | | | CLO | | | ELO | | |
|---|---|---|---|---|---|---|---|---|---|---|---|---|---|---|---|---|---|
| F | HA | N | N% | $\sigma_\alpha$ $\sigma_\delta$ | | N% | $\sigma_\alpha$ $\sigma_\delta$ | | N% | $\sigma_\alpha$ $\sigma_\delta$ | | N% | $\sigma_\alpha$ $\sigma_\delta$ | | N% | $\sigma_\alpha$ $\sigma_\delta$ | | M |
| | (hs) | | | (mas) | | | (mas) | | | (mas) | | | (mas) | | | (mas) | | |
| 1 | -0.8 | 138 | 71 | 26 | 28 | 87 | 19 | 21 | 83 | 18 | 20 | 85 | 18 | 21 | 81 | 18 | 21 | 2 |
| 2 | +3.0 | 39 | 46 | 18 | 22 | 64 | 26 | 28 | 64 | 27 | 30 | 53 | 27 | 25 | 53 | 28 | 18 | 2 |
| 3 | +2.9 | 45 | 82 | 22 | 27 | 87 | 18 | 15 | 87 | 19 | 15 | 87 | 18 | 16 | 87 | 20 | 15 | 3 |
| 4 | +3.0 | 14559 | 11 | 33 | 33 | 15 | 31 | 30 | 15 | 31 | 30 | 13 | 31 | 31 | 13 | 31 | 31 | 2 |
| 5 | +4.9 | 29944 | 94 | 587 | 549 | 82 | 315 | 303 | 85 | 380 | 355 | 82 | 381 | 366 | 85 | 450 | 430 | 8 |

*Notes*: F = FOV, HA = hour angle, N = number of Gaia stars on FOV, N% = percentage of used Gaia stars in reductions, $(\sigma_\alpha, \sigma_\alpha)$ = mean error of reductions, PGM = Photogravity Center Method, CGA = Circular Gaussian, EGA = Elliptical Gaussian, CLO = Circular Lorentzian, ELO = Elliptical Lorentzian, M = Model (Table 1). Outlier Gaia stars with |O-C| > 60 mas were eliminated in the reductions of FOVs 1–4. Outliers in the FOV 5 reductions were eliminated by sigma-clipping with a "relaxed" 5.0 factor.

we used the equivalent width $\sigma_E = \sqrt{ab}$, with $(a, b)$ being the fitted semi-major and semi-minor axes (Sections 4.4 and 4.5), and $FWHM = \sqrt{2ln2}\sigma_E$. Lorentzian widths were scaled to the Gaussian statistical width by dividing the Lorentzian semi-axes $(a, b)$ or $r_0$ by $\sqrt{2\beta ln2}$ (Section 4.5). Fig.12 shows that the FWHM (seeing) is nearly constant for all stars and PSFs, with a scatter growth toward fainter magnitudes. Fig.12 also shows that the PGM and PSF width distributions are distinct, with systematically smaller PGM values, except for the bright magnitude regime, and specially for magnitude < 14 in FOVs 4–5 where they become even larger than the PSF ones. For Gaussian profiles, the width from second moments (Section 3.6) grows and converges asymptotically for the statistical Gaussian width as the aperture radius increases. The optimal BOIA aperture radius is relatively smaller for fainter objects, resulting on smaller PGM widths. But BOIA apertures enlarge faster than the profile sizes as the object brightness grows and the PGM widths start to converge to Gaussian ones. For very bright stars, the profile is no longer Gaussian in the centre because the linear range of the detector is trespassed. Shallower central profiles result on broader moment-based widths in comparison with PSF ones. Larger radii also increase aperture flux contamination, specially on crowded stellar fields, resulting in larger second moments and again on broader PGM widths. This reasoning is also clearly supported by Fig.13 (Section 10.2.1). Since PSF fitting is less sensitive to these issues, the object's width is more accurately determined with Gaussian or Lorentzian profiles.

#### 10.2.1. BOIA explains and expands the 2–3 $\sigma$ aperture radius photometric and astrometric axiom

In the 70's and 80's, Auer, van Altenna, Chiu, Stetson and Stone pioneered astrometric works on digital images by the investigation of digitized plates and simulated images – see Stone (1989) and references therein, known as VAACS papers. From empirical experience, they recommend apertures with radius ranging between 2–3 times the Gaussian statistical width $\sigma$, or equivalently between 0.85–1.28 FWHM for better astrometric precision in the PSF fittings. From image simulations, a value of 1.25 FWHM was recommended by Stone (1989). These benchmark values are widely used up to date for setting the aperture radius in astrometric and photometric work.

Fig.13 displays for all FOVs the ratio between the BOIA aperture radius and the FWHM with magnitude for the PGM and each PSF. Benchmark 0.85 and 1.28 radius/FWHM ratio values are also plotted. For all FOVs and PSFs, the well sampled objects at intermediate magnitude ranges lay inside the recommended VAACS limits. Thus, BOIA's principle of setting aperture sizes by the best S/N gives a sound explanation for the old and long-standing astrometric/photometric axiom that recommends apertures of 2–3 $\sigma$ radius.

But Fig.13 tells an even broader history. BOIA's radius/FWHM ratios form a continuous pattern not only inside but also outside the 2–3 $\sigma$ limits, indicating how larger ratios must be for brighter objects and how smaller for fainter ones in a consistent way with magnitudes.

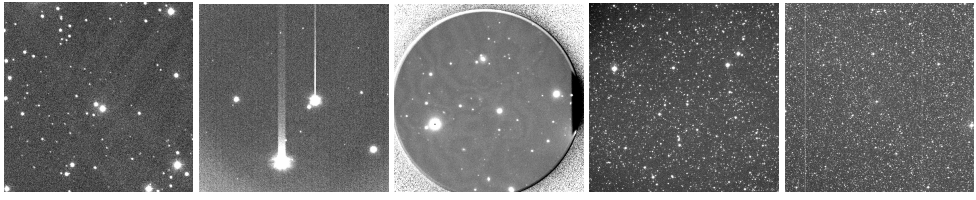

Figure 11: FOVs 1–5 from left to right. FOV 1 is a 60s exposure of 3'x3' taken in 2019 close to the meridian with the 1.6m telescope (IAU 874) at OPD Observatory with objects ranging from 11 to 21 G magnitudes. It is a starry field at the M35 cluster with some local empty regions without visible stars due to absorbing clouds. FOV 2 is a 60s exposure of 6'x6' sizes taken in 2017 far from the meridian with the OPD 1.6m telescope with objects ranging from 10.5 to 21 G magnitudes. It has few stars, some very faint, some very bright with CCD leaking features. The observation was made during a campaign to observe Neptune and Triton, visible in the field. FOV 3 is a 180s exposure taken in 2021 far from the meridian at the 4m SOAR telescope (IAU I33) with the Goodman Spectrograph Imager with an effective circular shape field of 7.2' radius. Objects range between 13 – 22 G magnitudes. It has very few stars per square arc minutes, and not as many bright ones as in FOV 2 – a fact compensated by the longer exposure. Notice the vignetting and the fringes in the sky background. The comet Bernadinelli–Bernstein is one of the faint objects visible near the FOV centre. FOV 4 is a 60s exposure of 12'x12' sizes taken in 2011 far from the meridian with the OPD 0.6m telescope (IAU 874) with objects ranging from 11 to 21 G magnitudes. It is an extremely crowded star field near the Galactic Center. The image was taken in an astrometric campaign to observe the Pluto/Charon system. FOV 5 is a large and starry field with poor pixel resolution (pixel scale = 1.079 "/pixel), with objects ranging between 11 and 21 G magnitudes. The 170s observation extremely low in the horizon was made at the Rozhen National Astronomical Observatory of Bulgaria (IAU 71) for the astrometric campaign of the PHA Apophis during its 2013 apparition.

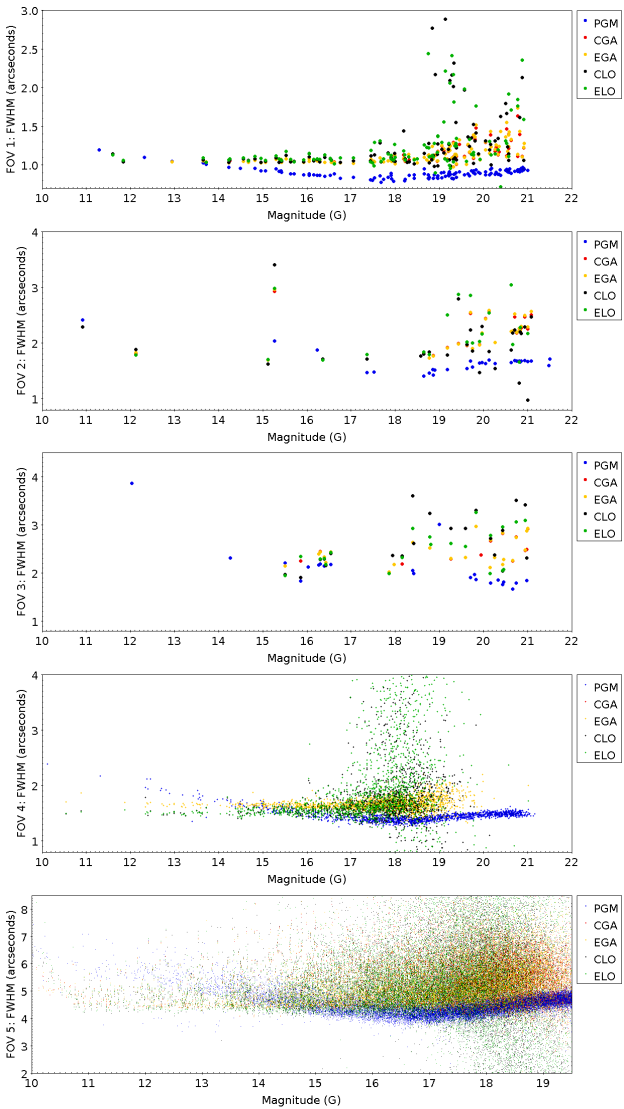

Figure 12: FWHM as a function of magnitude for PGM (blue), for Circular (red) and Elliptical (yellow) Gaussian, Circular (black) and Elliptical (green) Lorentzian PSFs, for the FOVs described in Section 10.1.

BOIA optimizes the radius/FWHM ratio for already well sampled objects as opposed to the common practice of predefining apertures with 2–3 $\sigma$ sizes. BOIA also naturally finds the most adequate radius/FWHM ratio in bad sampling conditions – saturation or under-exposition. BOIA represents an improvement over the current practice of setting aperture sizes for virtually any magnitude and object's sampling conditions in astrometric and photometric work.

### 10.3. The (x,y) errors of PSF fittings under BOIA apertures

Mighell (2005) estimates *lower limits* for the centroid precision of Circular Gaussian profile fittings (see also Camargo et al. 2022). From Mighell's equations 47 and 48 for bright and faint source regimes, we computed expected *lower limits* for the total $(x, y)$ errors of our Circular Gaussian fittings of FOVs 1–5 by the use of the formulae in Eq. 19. Here, $\sigma$ is the statistical Gaussian width, $\beta$ is the Gaussian's *effective background area* with highness $h$ expressed in Gaussian volume $V$ by $h = V/(2\pi\sigma^2)$ (Section 2.1 in Mighell 2005), $B$ and $F$ are the error contributions in the bright and faint source regimes, $NBG$ is the number of sky background ring pixels used in the sky background estimation, $Error_x = Error_y$ the individual $x$ and $y$ errors and $Error(x, y)$ the total error of the $(x, y)$ centre.

$$\beta = \frac{4\pi\sigma^2}{(gain.flux)^2} \;,\; B = \frac{1}{S/N} \;,\; F = B\frac{\sqrt{2}}{1+\sqrt{\frac{\beta}{NBG}}}$$

$$E_x = E_y = \sigma\sqrt{B^2 + F^2} \;,\; Error\,(x,y) = \sqrt{E_x^2 + E_y^2} \quad (19)$$

We compare the *lower limit* errors from Eq. 19 with the errors from the Circular Gaussian fittings of FOVs 1–5 in Fig.14. The $(x, y)$ precision of our Circular Gaussian fittings is close to the lower limit, i.e. BOIA-based $(x, y)$ measurements do extract the most of astrometric information out of the data for all FOVs.

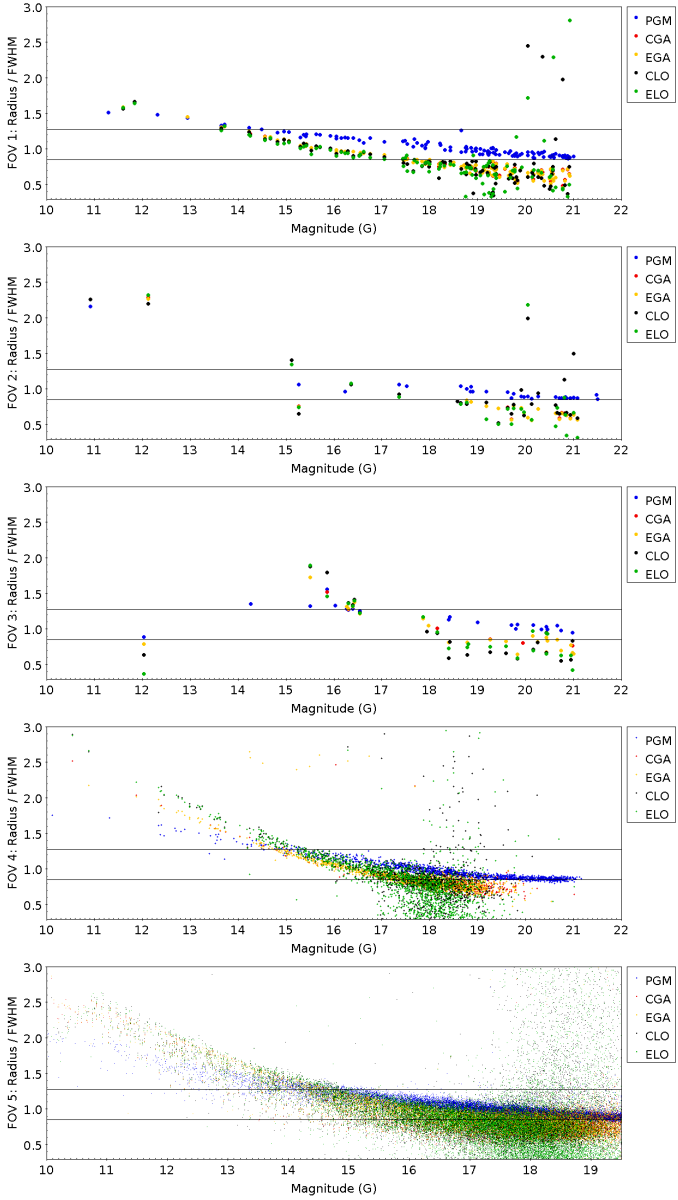

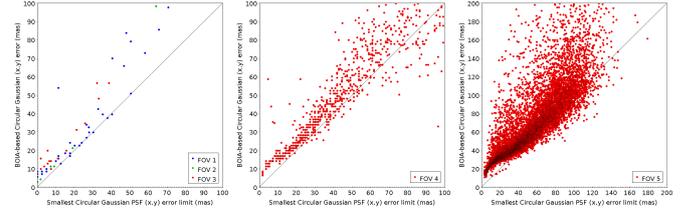

Figure 14: Lower limit $(x, y)$ errors computed by Eq.19 compared to the errors from actual Circular Gaussian fittings of FOVs 1–5.

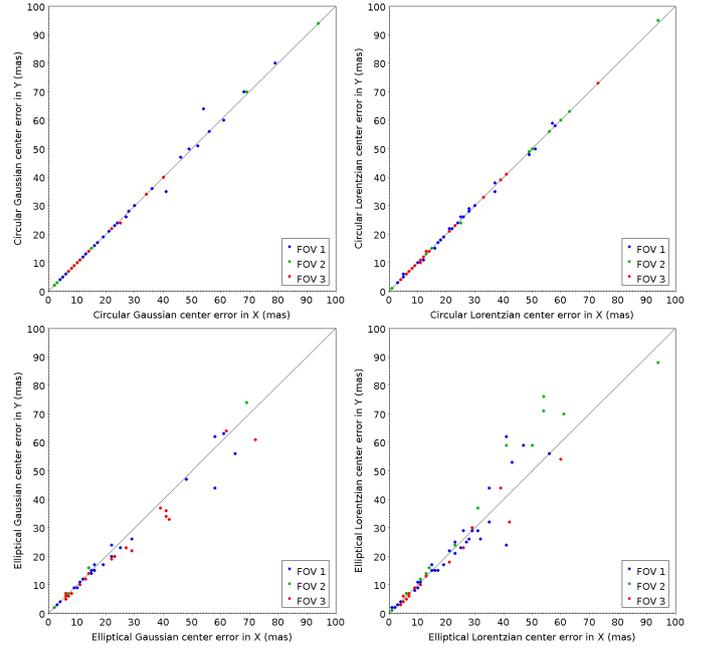

Figure 15: Measured $(x, y)$ centre errors along right ascension ($X$) compared with the ones along declination ($Y$) from PSF fittings of Circular and Elliptical Gaussians (left panels) and Lorentzians (right panels) for FOVs 1–3.

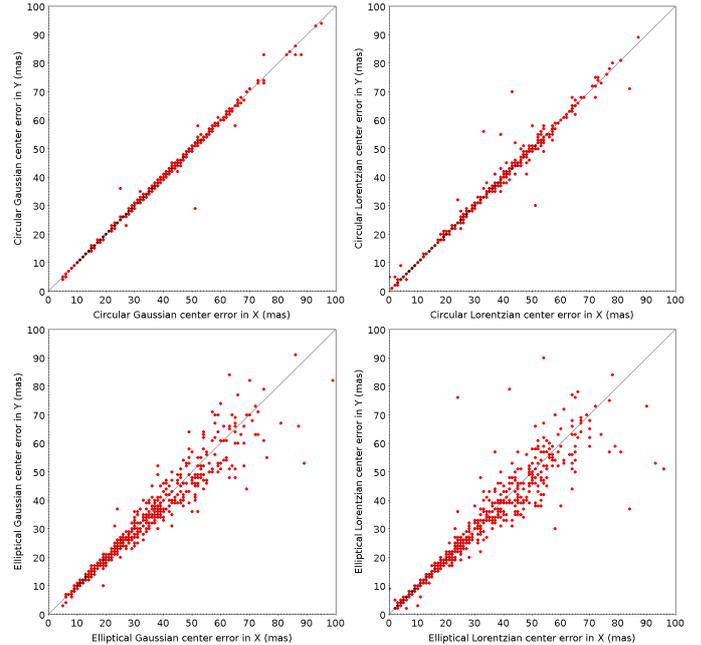

Figure 13: `BOIA`'s aperture radius/FWHM ratios against magnitudes for PGM (blue), for Circular (red) and Elliptical (yellow) Gaussian, Circular (black) and Elliptical (green) Lorentzian PSFs, for all FOVs (Section 10.1). Lines indicate apertures with radius between 2–3 $\sigma$, or 0.85–1.28 FWHM.

The $(x, y)$ errors are translated to $(\alpha, \delta)$ errors (see Section 8.3), so that we can properly evaluate measurement error features (if any) in terms of right ascension and declination. Fig.15, 16 and 17 display $X$ versus $Y$ errors ($\alpha$ .vs. $\delta$ measuring errors) for FOVs 1–3, 4 and 5 respectively. In all plots, the $X$ errors are remarkably equivalent to the $Y$ ones, with less dispersion for the circular PSFs. The circular PSFs fitted the pixel data better than elliptical ones, suggesting no image deformation along right ascensions exists, like bad sidereal guiding. No significant color refraction is present too, which is expected since the I filter was used in FOVs 1–4, and the R filter in FOV 5, in spite of the large hour angle in most observations (see Table 3 in Section 10.1).

Figure 16: Same as in Fig.15 for FOV 4.

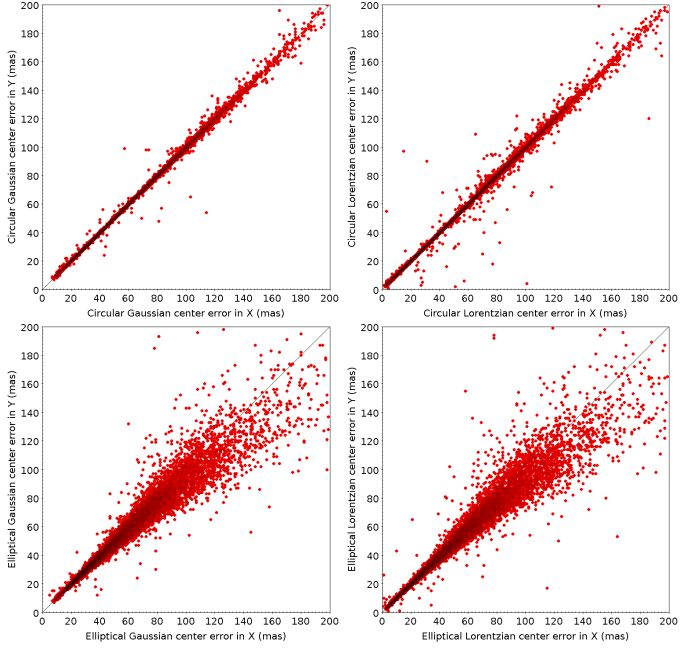

Figure 17: Same as in Fig.15 and 15 for FOV 5.

## 10.4. PGM performance compared with PSF fittings

Fig.18, 19 and 20 show the typical $(x, y)$ error degradation with increasing magnitudes for all PSF fittings and PGM centres in FOVs 1–3, 4 and 5 respectively. Fig.21 displays $X$ .vs. $Y$ centre errors ($\alpha$ .vs. $\delta$ errors) computed with Eq. 8 (Section 4.2) for PGM measurements made for all FOVs. The self-consistency of PGM $(x, y)$ errors is the same found for fitted PSF centres (Section 10.3).

In general, we do not expect that the PGM (or any other Moment method) have the same performance of PSF fittings for all error ranges and FOVs (see Stone 1989). This depends on *seeing*, sky background luminosity and object brightness. Nevertheless, we verify that the behaviour of PGM centre errors computed by Eq. 8 in comparison with PSF ones are remarkably consistent with what we would expect, based on the FOV characteristics and $(\alpha,\delta)$ reductions displayed in Table 2 and Table 3. This comparison is shown in Fig.22, Fig.23 and Fig.24 for FOVs 1–3, FOV 4 and FOV 5 respectively. Comparisons are made for errors on each $X$ (or $\alpha$) and $Y$ (or $\delta$) coordinate, and for total errors $(E_\alpha^2 + E_\delta^2)^{1/2}$ for all PSFs. The scatter grows toward larger errors in all plots because: a) larger centre errors indicate deviations from the PSF model; b) the second moments of the object's image – on which Eq. 8 parameters are based – no longer characterize well the object's profile.

FOVs 1–3 correspond to the case of long focus, larger aperture telescopes with small FOVs and good pixel scale resolution/PSF sampling. FOV2 and FOV3 have few stars – less than 1 star per squared arc minute. FOV1 is a moderate starry field with ≈ 16 stars per squared arc minute. Fig.22 shows good agreement between PGM and PSF errors for brighter and fainter objects (small and large error ranges), and a slightly better performance for all PSFs in between, as expected for these kinds of FOVs.

FOV 4 and FOV 5 come from smaller 50–60 cm aperture telescopes, with mild (FOV 4) and poor (FOV 5) pixel scale resolutions, larger (FOV 4) and very large (FOV 5) sizes. FOV 4 is crowded with stars. FOV 5 behaves like a crowded star field due to the poor pixel scale and moderate star density (9 per squared arc minute). In these cases, PGM performance is systematically inferior to PSF's, due to contamination by nearby sources. PGM and PSF centre errors are more comparable at the very small error range in all plots where objects are much brighter and contamination is less relevant. Notice the better performance of Lorentzian profiles with respect to Gaussian ones at this very bright range, since non-linear photo-counting and saturated profiles are better handled by these PSFs.

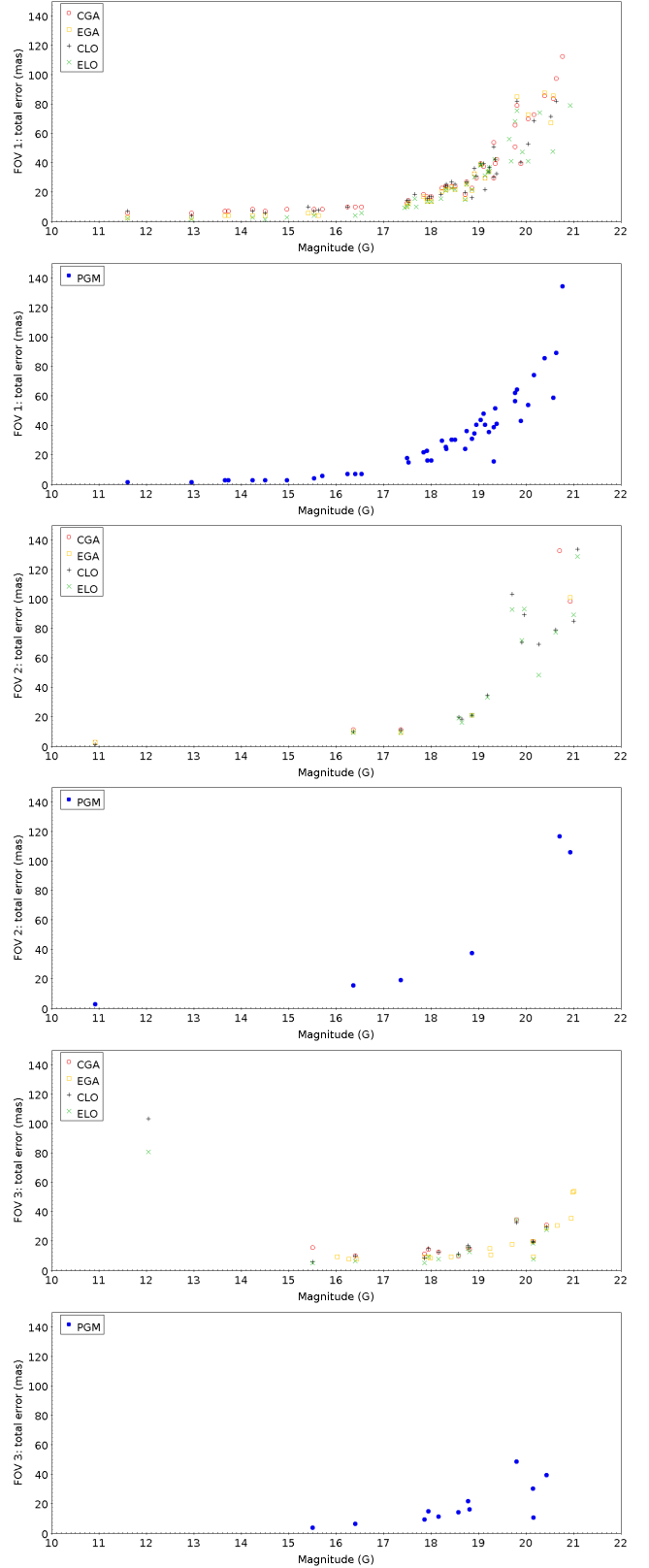

Figure 18: Total $(x, y)$ centre error with magnitudes for Circular and Elliptical Gaussian and Lorentzian PSFs, and for PGM for FOVs 1–3.

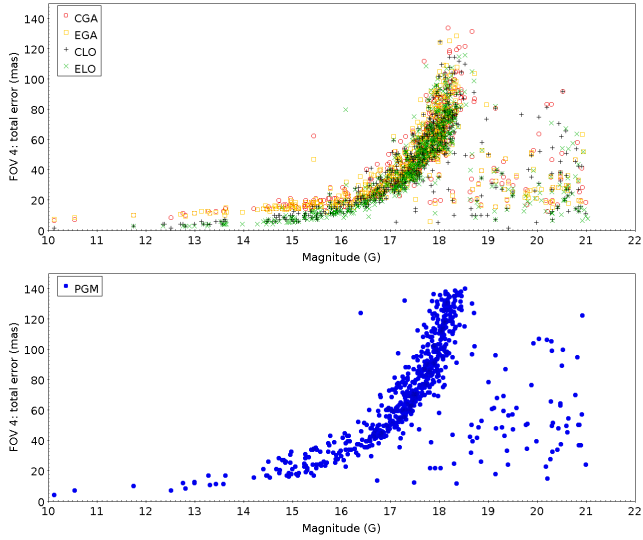

Figure 19: Same as in Fig. 18 for FOV 4.

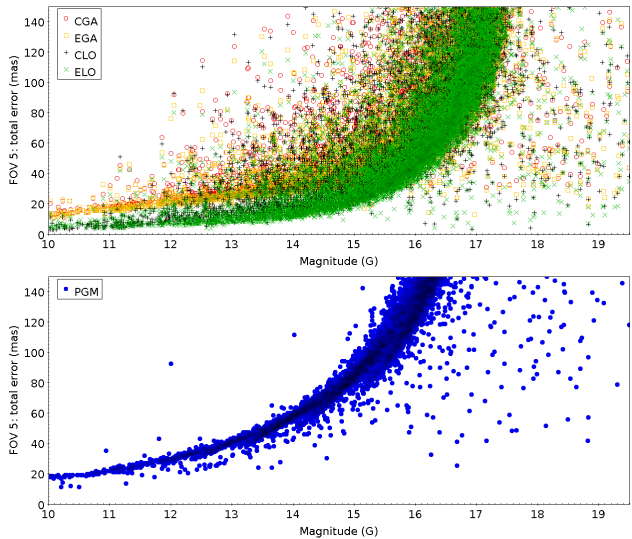

Figure 20: Same as in Fig. 18 and Fig.19 for FOV 5.

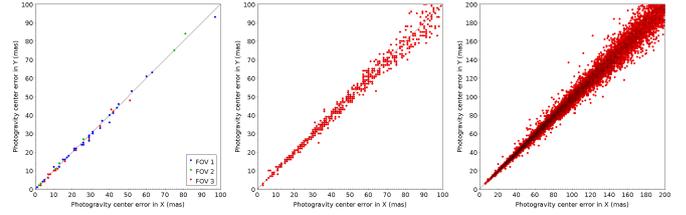

Figure 21: PGM measured $(x, y)$ centre errors along right ascension ($X$ errors) against declination ones ($Y$ errors) obtained with Eq.8 for all FOVs.

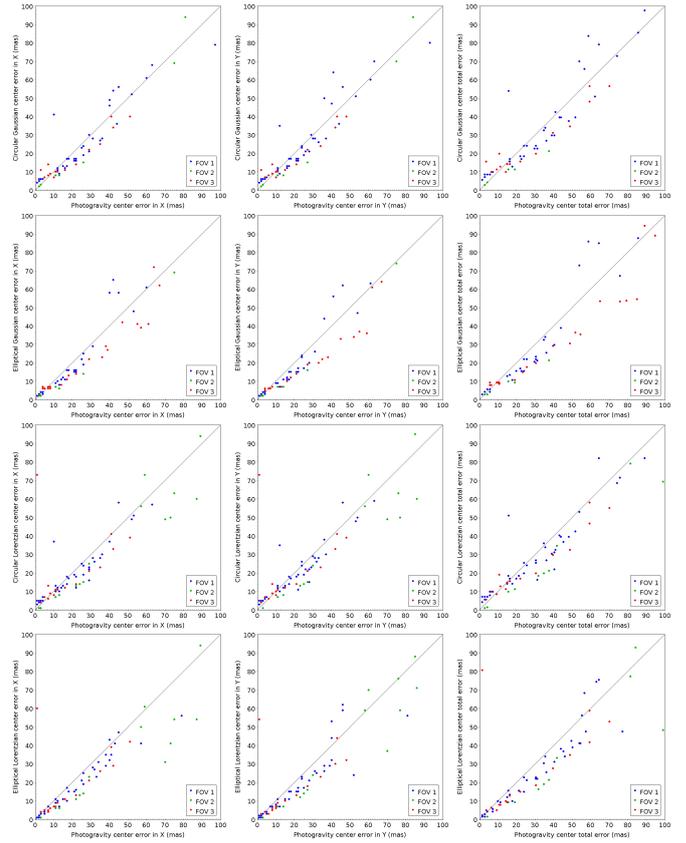

Figure 22: Comparison of PGM measured $(x, y)$ centre errors along right ascension ($X$ errors) and declination ($Y$ errors) with those from Circular and Elliptical Gaussians (half top panels) and Circular and Elliptical Lorentzians (half lower panels) for FOVs 1–3.

## 11. PAT validation: published works

PAT has been used by the Rio Group since the 1990's in works involving reference systems, natural satellites' and NEAs' astrometry for dynamical and ephemeris studies, and within the international collaboration under the Lucky Star umbrella since 2006 for the precise prediction of stellar occultations by planetary satellites, dwarf-planets, TNOs, Centaurs and Trojan asteroids. The reader will find practical usage and certification of PAT in all 1992–2022 referred works listed in Table 4. It gives the used task version available at the epoch of observations (see Section 2), the type of study, the object class and name, the journal and the reference of the publication with the year. Sometimes the use of the PRAIA package is not explicitly stated in the text, and sometimes old and quite generic references (Assafin, 2006; Assafin et al., 2011) are given.

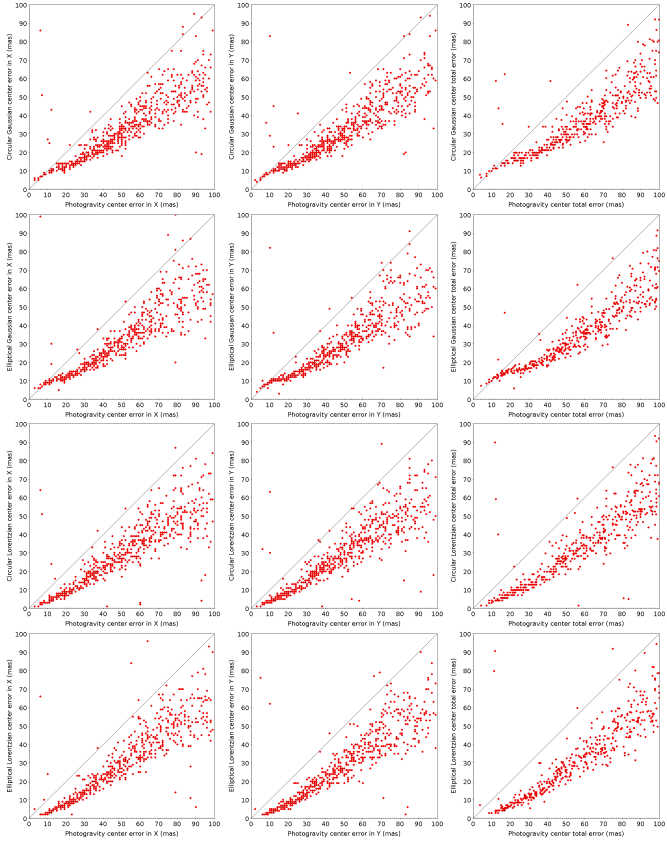

Figure 23: Same as in Fig. 22 for FOV 4.

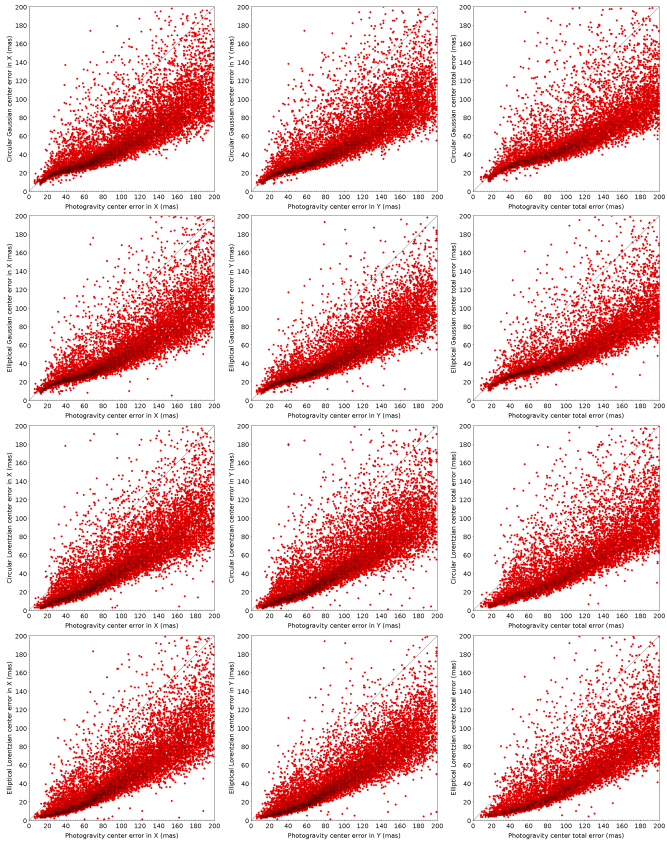

Figure 24: Same as in Fig. 22 for FOV 5.

Table 4: 53 referred works (1992–2022) that used PRAIA astrometry task.

| V | T | C | Body | Journal & Reference |
|---|---|---|------|---------------------|
| 0 | R | s | Stars | PASP: Vieira, Assafin & Martins (1992) |
| 0 | R | Q | Quasars | A&AS: Assafin & Martins (1992) |
| 0 | R | R | Radio stars | A&AS: Assafin et al. (1996) |
| 0 | S | S | Nereid | A&AS: Veiga et al. (1996) |
| 0 | R | sQ | Stars, Quasars | AJ: Assafin, Martins & Andrei (1997) |
| 0 | R | sQ | Stars, Quasars | AJ: Assafin et al. (1997) |
| 0 | R | R | Radio stars | AJ: Andrei et al. (1999) |
| 0 | R | Q | Quasars | AJ: da Silva Neto et al. (2000) |
| 1 | R | Q | Quasars | AJ: Assafin et al. (2003) |
| 1 | R | Q | Quasars | A&A: da Silva Neto et al. (2005) |
| 1 | R | Q | Quasars | AJ: Assafin et al. (2005) |
| 1 | O | S | Charon | Nature: Sicardy et al. (2006) |
| 1 | R | Q | Quasars | A&A: Assafin et al. (2007) |
| 2 | O | KSs | Pluto system | A&A: Assafin et al. (2010) |
| 2 | O | S | Charon | AJ: Sicardy et al. (2011) |
| 2 | R | Q | Quasars | A&A: Camargo et al. (2011) |
| 2 | O | D | Eris | Nature: Sicardy et al. (2011) |
| 2 | O | Ks | TNOs | A&A: Assafin et al. (2012) |
| 2 | A | A | (99942) Apophis | A&A: Bancelin et al. (2012) |
| 2 | O | D | Makemake | Nature: Ortiz et al. (2012) |
| 2 | R | Q | Quasars | MNRAS: Assafin et al. (2013) |
| 2 | O | K | (50000) Quaoar | ApJ: Braga-Ribas et al. (2013) |
| 2 | O | Ks | TNOs | A&A: Camargo et al. (2014) |
| 2 | O | D | Pluto | A&A: Boissel et al. (2014) |
| 2 | O | C | Chariklo | Nature: Braga-Ribas et al. (2014) |
| 2 | O | D | Pluto | A&A: Benedetti-Rossi et al. (2014) |
| 2 | O | D | Pluto | Icarus: Olkin et al. (2015) |
| 2 | SO | S | Irregular satellites* | A&A: Gomes-Júnior et al. (2015) |
| 2 | A | S | Uranus satellites | A&A: Camargo et al. (2015) |
| 2 | O | D | Pluto | ApJ: Dias-Oliveira et al. (2015) |
| 2 | A | A | (99942) Apophis | A&A: Thuillot et al. (2015) |
| 2 | O | K | TNOs | A&A: Desmars et al. (2015) |
| 2 | O | D | Pluto | ApJL: Sicardy et al. (2016) |
| 2 | SO | S | Irregular satellites* | MNRAS: Gomes-Júnior et al. (2016) |
| 2 | O | K | 2007 $UK_{126}$ | AJ: Benedetti-Rossi et al. (2016) |
| 2 | O | K | 2003 $AZ_{84}$ | AJ: Dias-Oliveira et al. (2017) |
| 2 | O | C | Chariklo | AJ: Bérard et al. (2017) |
| 2 | O | C | Chariklo | AJ: Leiva et al. (2017) |
| 2 | O | D | Haumea | Nature: Ortiz et al. (2017) |
| 2 | O | K | TNOs | P&SS: Camargo et al. (2018) |
| 3 | O | K | TNOs | AJ: Banda-Huarca et al. (2019) |
| 3 | O | D | Pluto | A&A: Desmars et al. (2019) |
| 3 | O | K | 2003 $VS_2$ | AJ: Benedetti-Rossi et al. (2019) |
| 3 | O | S | Phoebe (S IX) | MNRAS: Gomes-Júnior et al. (2020) |
| 3 | O | K | 2002 $TC_{302}$ | A&A: Ortiz et al. (2020) |
| 3 | O | K | Varda | A&A: Souami et al. (2020) |
| 3 | O | KC | TNOs, Centaurs | A&A: Rommel et al. (2020) |
| 3 | O | K | 2002 $GZ_{32}$ | MNRAS: Santos-Sanz et al. (2021) |
| 3 | O | C | Chariklo | A&A: Morgado et al. (2021) |
| 3 | O | D | Pluto | ApJL: Sicardy et al. (2021) |
| 3 | S | S | Uranus satellites | P&SS: Camargo et al. (2022) |
| 3 | O | S | Triton | A&A: Marques Oliveira et al. (2022) |
| 3 | O | K | 2003 $VS_2$ | A&A: Vara-Lubiano et al. (2022) |
| 3 | O | K | Huya | A&A: Santos-Sanz et al. (2022) |

*Note*: Task version (V): (Section 2). Study type (T): reference systems (R), natural satellites (S), asteroids (A), stellar occultation prediction (O). Body class (C): Dwarf planet (D), Satellite (S), Asteroid (A), Trojan (T), Kuiper Belt Object (K), Centaur (C), Quasars (Q), Radio stars (R), Stars (s). (*): Jupiter irregular satellites, Phoebe (Saturn).

## 12. Supplementary PRAIA astrometry tasks

We developed and implemented some specific astrometric procedures as supplementary `PRAIA` astrometry tasks.

Mutual approximations is a new astrometric technique first introduced in Morgado et al. (2016) for Galilean satellites. It measures central instants at the closest approach between two moving satellites in the sky plane. Measurements are made on small portions of the FOV, benefiting from the *precision premium* (Lin et al., 2019). Approximations share the same geometric principles and parameters of mutual phenomena, with similar precision and the advantage that they always occur, while mutual phenomena can only be observed at the planet's equinoxes. Also, central instants do not depend on reference stars and are useful in orbit and ephemeris fittings. The technique was later further improved and also applied to Uranus' satellites (Morgado et al., 2019; Santos-Filho et al., 2019).

A new astrometric technique called *Differential Distance Astrometry* was also implemented as another `PRAIA` task. Distance ratios between pairs of moving objects are measured, serving as observables for ephemeris fittings. A paper describing the method with results for the Uranus system is in preparation (Assafin et al., 2023).

Another useful supplementary `PRAIA` task allows for the automatic extraction of ICRS astrometric topocentric and airless apparent ephemeris for Solar System bodies through the internet, using the JPL SPICE/NAIF toolkit[8]. The task is a flexible tool used for constructing Ephemeris catalogues (Section 5), or for preparing the target lists of Solar System bodies (Section 6).

## 13. PAT and popular astrometric packages

Only a handful of published astrometry programs are at `PAT`'s niche to the best of our knowledge. Some are used for specific tasks like object detection within scripts. Worth mentioning are `SExtractor` (Bertin & Arnouts, 1996), `SCAMP`[9] (Bertin, 2006), `ASTROMETRY.NET`[10] (Lang et al., 2010), `ASTROMETRICA`[11] and `AAPPDI` (Zhang et al. 2022, and references therein).

`SExtractor` combines sigma-clipping and histogram mode to set local sky background thresholds for object detection. It derives $(x, y)$s from a multiple isophotal analysis technique, useful for fast source deblending in crowded star fields at the cost of centroid accuracy.

Starting from `SExtractor` detections, `SCAMP` does astrometry if World Coordinate System (WCS) data are accurate within a few arcminutes and some parameters are given, as the approximate pixel scale. One dimension weighted Gaussian fittings give $(x, y)$s for fast processing. Pixel scale and image orientation are found by the analysis of 2D histograms of source pair coordinates in the log(projected distance) - position-angle space for measurements and reference catalogue tangent plane coordinates. An overlapping method (Eichhorn, 1960) binding many observations of the same FOV is possible.

`ASTROMETRY.NET` automatically finds the pointing, scale, and orientation of an input image without any a priory WCS knowledge. For accomplishing these goals with least computer time consumption four to five "well imaged" stars suffice. `ASTROMETRY.NET` is not designed for robust detection of all sources in the FOV nor high precision centring. Detection is made after flattening the sky background by a fast median-filter and by finding bright grouped pixels. Centres are determined by fitting Gaussian profiles over $3x3$ pixel grids centred at the peak. If the centre falls outside the grid it is reset to the central pixel of the grid. Coding rectangles by four stars on different scales and sky locations on a reference catalogue and hierarchically searching for matches with the same coding on the measured detections yield the identification of reference catalogue stars in the FOV, enabling the estimation of the pointing, scale, and orientation angle. `ASTROMETRY.NET` can only robustly identify objects in non-distorted FOVs, solely affected by scale and rotation – FOV identification may even fail under significant distortions.

`ASTROMETRICA` is a popular shareware (paid) program used by the amateur community for asteroid astrometry. It has a rich GUI with a series of useful operations on asteroid observation, including interfaces with the Minor Planet Center and Vizier, and image stacking. It is designed for human intervention and is heavily interactive in an image-by-image basis. FOV parameters must be furnished by the WCS or the user, with $(x, y)$ measurements and FOV identification performed in semi-automatic way. No manual or reference paper are available, making it difficult to understand what kind of PSF fitting is performed for centring (presumably a 2D Gaussian profile), what tangent plane models and astrometric reduction procedures are available, and how errors are computed.

`AAPPDI` (Automatic Astrometric Processing Program for Digital Images) is a private `FORTRAN` software developed and used by a group of Chinese researchers, recently capable of automatic object detection, centring, reference star matching, $(\alpha, \delta)$ reduction, star and target position determination. For detection, a median filter is used and a 1-sigma threshold above the local sky background certifies pixels to be used in the centring of objects. Centring is based on Modified Moment methods, one of which uses the square of counts as pixel coordinate weights. Reference star matching is made with prior knowledge of pixel scale, assuming that the tangent point $(\alpha_0, \delta_0)$ coordinates are precisely known. A linear model is used and small angle rotations applied to the projected catalogue coordinates until they match the star measurements. Polynomials of high order can be used in the tangent plane fittings. Comparison with ephemeris allow for the identification of measured target $(\alpha, \delta)$ positions.

Our astrometry demands high accuracy/precision for a large number of observations from heterogeneous telescopes not rarely lacking instrumental information. That is why the $(x, y)$s from all these softwares are sub-optimal with respect to `PAT`. The `BOIA` method is intrinsically more efficient in the detection of objects, particularly in the limit of noise, than `SExtractor`, `SCAMP`, `ASTROMETRY.NET`, and even `AAPPDI`. `PAT` does that virtually without any dependence on parameters or sky-noise factors, for a wide variety of sky background conditions, including strong vignetting. `PAT` is fully automatic from object detection to reference catalogue identification (no a priory pixel scale information is needed) to $(\alpha, \delta)$ reduction. `PAT`'s PSFs and PGM centring algorithms perform better – see papers in Table 4 and references therein.

---

[8] JPL SPICE/NAIF toolkit: https://naif.jpl.nasa.gov/naif/toolkit.html
[9] SCAMP: https://www.astromatic.net/software/scamp/
[10] ASTROMETRY.NET: https://astrometry.net
[11] ASTROMETRICA: http://www.astrometrica.at/

Although `SCAMP` presents interesting features, it lost support since 2014 and has no modern reference catalogues available such as the Gaia DR3. `SCAMP` only works if the WCS information is accurate enough, including knowledge of the pixel scale.

`ASTROMETRY.NET` is not a genuine astrometric tool since it does not perform actual reductions in the tangent plane. It just identifies objects and common catalogue stars with measured $(x, y)$s that match their tangent plane coordinates by simple scaling and rotation, ignoring possible FOV distortions. On the other hand, `PAT` does complex tasks from object detection to $(\alpha, \delta)$ reduction in a very fast way. `ASTROMETRY.NET` takes about 5 – 30 s to determine WCS parameters for FOVs ranging from 512 x 512 to 4096 x 4096 pixels, the same amount of time taken by `PAT` to fully complete the astrometry of these FOVs with a Intel® Core™ i7-6700HQ CPU at 2.60GHz.

Unlike `PAT`, `ASTROMETRICA` is not designed for dealing with huge amounts of images. The lack of documentation and scientific referred publications supporting and validating the tool is also a drawback. Occasional tests with `ASTROMETRICA` in our collaboration work with the amateur community had `PAT` giving systematically better results.

Although promising, `AAPPDI` still lacks detailed documentation and is parameter-dependent on most modules, like in object detection and reference catalogue star identification. Also, only Modified Moment centring algorithms are available, all subject to the problems overcome by the PGM method (see Section 4.1).

Although not appealing from a GUI aesthetic point of view, `PAT` may become a very attractive astrometric tool for the amateur community in the future because of the simplicity in the usage and fast performance, even for a great amount of heterogeneous observations. Besides, it is an astrometry tool certified by specialists from the professional astronomical community.

## 14. Conclusions, remarks, future updates

We presented the astrometry concepts behind the Package for the Reduction of Astronomical Images Automatically – `PRAIA`. Differential aperture photometry with `PRAIA` is described in detail on another paper (Assafin, 2023a), as well as digital coronagraphy (Assafin, 2023b). `PRAIA` gives photometry/astrometry support for the studies carried out by the Rio Group and the international collaboration under the Lucky Star Project. The standalone codes have no dependencies with external packages and are written in `FORTRAN`. They are simple to install and use. The package does not suffer from deprecation issues as is common with Python-based packages. Unlike packages such as `IRAF` or `SCAMP` which lost support, `PRAIA` has constant updates with new releases and, from time to time, upgrades to new versions since 1992 (astrometry) and 2006 (photometry).

The `PRAIA` astrometry task (`PAT`) does accurate/precise fast automatic astrometry on FOVs. `PAT` has no size limits to measure sources, with PSFs more effective in the astrometry of point-like to moderately extended sources, although in principle one can do fine by using `PAT`'s Lorentzian models for extended profiles. `PAT` is suited for the $(\alpha, \delta)$ reduction of huge amounts of observations from heterogeneous instruments for reference system works, natural satellite and NEA astrometry for dynamical/ephemeris studies, and lately for the precise prediction of stellar occultations by planetary satellites, dwarf-planets, TNOs, Centaurs and Trojan asteroids. It has been used in 53 scientific publications between 1992–2022 (Table 4). Many astrometry innovations were incorporated to `PAT`, most of them related to the automatic detection/analysis of objects by `BOIA` (Section 3). `BOIA` explains and expands the old astrometry/photometry axiom that recommends the use of 2–3 $\sigma$ radius apertures (Section 10.2.1). A novel moment-based centring used by `PAT` – the Photogravity Center Method (PGM) – surpasses the Modified Moment by Stone (1989) and has $(x, y)$ error estimates (Sections 4.1 and 4.2).

`PAT` is superior in all typical (sometimes very difficult) observation conditions in comparison to astrometry tools and packages such as `SExtractor`, `SCAMP`, `ASTROMETRY.NET`, `ASTROMETRICA` and `AAPPDI` – the most used packages at `PAT`'s niche to the best of our knowledge.

Besides works within the Rio Group and Lucky Star collaboration, `PAT` was officially assigned as the astrometric supporting tool for the activities of participants in the Gaia Follow-Up Network for Solar System Objects (GAIA-FUN-SSO[12], Thuillot 2011; Assafin et al. 2011). `PAT` is also being used in the astrometry and occultation predictions to TNOs and Centaurs observed within the Dark Energy Survey (DES, see Banda-Huarca et al. 2019 and references therein). Besides Solar System work, `PAT` can also be used in any astrophysical work context.

Future updates include the use of elliptical apertures, memory optimization and parallelization for improving speed. `PAT` runs on MacBook and debian Linux operation systems. It was thoroughly tested with Ubuntu distributions 16.04–22.04. We will encapsulate `PAT` in docks to use it on any Linux/Windows systems and run it as an astrometry service to the astronomical community in the LIneA Solar System Portal[13]. `PAT` code, input files and documentation are publicly available for the first time at https://ov.ufrj.br/en/PRAIA/.

## Declaration of competing interest

The author declares not having any known competing financial interests or personal relationships that could have appeared to influence the work reported in this paper.

## Acknowledgements

We thank Dr. J. I. B. Camargo, Dr. R. Vieira Martins, Dr. A. R. Gomes Júnior, Dr. B. E. Morgado, Dr. G.

---

[12]GAIA-FUN-SSO: https://gaiafunsso.imcce.fr/
[13]LIneA stands for Interinstitutional Laboratory for e-Astronomy. LIneA Solar System Portal: https://tno-dev.linea.org.br/

Benedetti Rossi, Dr. S. Santos Filho and J. Arcas Silva for the fruitful discussions in some aspects of the task upgrades and for helping in the internal revision of the text. M.A. thanks CNPq grants 427700/2018-3, 310683/2017-3 and 473002/2013-2. The Lucky Star Project agglomerates the efforts of the Paris, Granada and Rio teams, and is funded by the European Research Council under the European Community's H2020 (ERC Grant Agreement No. 669416). This work has made use of data from the European Space Agency (ESA) mission *Gaia* (https://www.cosmos.esa.int/gaia), processed by the *Gaia* Data Processing and Analysis Consortium (DPAC, https://www.cosmos.esa.int/web/gaia/dpac/consortium). Funding for the DPAC has been provided by national institutions, in particular the institutions participating in the *Gaia* Multilateral Agreement.